\documentclass[12pt, draftclsnofoot, onecolumn]{IEEEtran}
\usepackage{amsmath,amsthm,amsfonts,amscd,amssymb,graphics,dsfont}
\usepackage[mathscr]{eucal}
\usepackage{graphics,graphicx,multicol}
\usepackage{epsfig}
\usepackage{epstopdf}
\usepackage{subfigure}
\usepackage{stfloats}
\usepackage{latexsym}
\usepackage{amsfonts}
\usepackage{cite}
\usepackage{algorithm,algorithmic}
\usepackage{color}
\usepackage{xcolor}
\usepackage{multirow}
\usepackage{booktabs}
\usepackage{diagbox}

\newcommand{\qqed}{\hfill $\blacksquare$}

\begin{document}
\title{A Bipartite Graph Neural Network Approach for Scalable Beamforming Optimization} 
\author{Junbeom Kim,~\IEEEmembership{Student Member,~IEEE}, Hoon Lee,~\IEEEmembership{Member,~IEEE}, \\ Seung-Eun Hong, and Seok-Hwan Park,~\IEEEmembership{Member,~IEEE} \vspace{-10mm}
\thanks{This work was supported in part by Institute of Information \& Communications Technology Planning \& Evaluation (IITP) Grants funded by the MSIT (No. 2021-0-00467, Intelligent 6G Wireless Access System) and (No. 2018-0-01659, 5G Open Intelligence-Defined RAN (ID-RAN) Technique Based on 5G New Radio), in part by the National Research Foundation (NRF) of Korea Grants funded by the MOE under Grant 2019R1A6A1A09031717 and by the MSIT under Grant 2021R1C1C1006557 and Grant 2021R1I1A3054575, and in part by BK21 FOUR Program by Jeonbuk National University Research Grant. \textit{(Corresponding author: Hoon Lee; Seok-Hwan Park.)}

J. Kim and S.-H. Park are with the Division of Electronic Engineering, Jeonbuk National University, Jeonju 54896, South Korea, and also with the Future Semiconductor Convergence Technology Research Center, Jeonbuk National University, Jeonju 54896, South Korea (e-mail: junbeom@jbnu.ac.kr; seokhwan@jbnu.ac.kr).

H. Lee is with the Department of Information and Communication Engineering, Pukyong National University, Busan 48513, South Korea (e-mail: hlee@pknu.ac.kr).

S.-E. Hong is with the Future Mobile Communication Research Division, Electronics and Telecommunications Research Institute, Daejeon 34129, South Korea (e-mail: iptvguru@etri.re.kr).

© 2022 IEEE.  Personal use of this material is permitted.  Permission from IEEE must be obtained for all other uses, in any current or future media, including reprinting/republishing this material for advertising or promotional purposes, creating new collective works, for resale or redistribution to servers or lists, or reuse of any copyrighted component of this work in other works.}}

\maketitle

\begin{abstract}
Deep learning (DL) techniques have been intensively studied for the optimization of multi-user multiple-input single-output (MU-MISO) downlink systems owing to the capability of handling nonconvex formulations. However, the fixed computation structure of existing deep neural networks (DNNs) lacks flexibility with respect to the system size, i.e., the number of antennas or users. This paper develops a bipartite graph neural network (BGNN) framework, a scalable DL solution designed for multi-antenna beamforming optimization. The MU-MISO system is first characterized by a bipartite graph where two disjoint vertex sets, each of which consists of transmit antennas and users, are connected via pairwise edges. These vertex interconnection states are modeled by channel fading coefficients. Thus, a generic beamforming optimization process is interpreted as a computation task over a weight bipartite graph. This approach partitions the beamforming optimization procedure into multiple suboperations dedicated to individual antenna vertices and user vertices. Separated vertex operations lead to scalable beamforming calculations that are invariant to the system size.
The vertex operations are realized by a group of DNN modules that collectively form the BGNN architecture. Identical DNNs are reused at all antennas and users so that the resultant learning structure becomes flexible to the network size. Component DNNs of the BGNN are trained jointly over numerous MU-MISO configurations with randomly varying network sizes. As a result, the trained BGNN can be universally applied to arbitrary MU-MISO systems. Numerical results validate the advantages of the BGNN framework over conventional methods.
\end{abstract}


\section{Introduction}
Beamforming techniques have been regarded as promising solutions for multi-user multi-antenna communication systems \cite{Bjornson:SPM14, Shi:TSP11,Christensen:TWC08,Schubert:TVT04, Boche:VTC02}.
The weighted minimum mean squared error (WMMSE) algorithm obtains a locally optimum beamforming solution for the weighted sum rate maximization \cite{Bjornson:SPM14, Shi:TSP11, Christensen:TWC08}. The fairness among multiple users is secured by maximizing the minimum rate performance, whose solution can be found by the convex/nonconvex optimization techniques \cite{Schubert:TVT04, Boche:VTC02}. The nonconvexity of the objective and the constraint functions, together with coupled beamforming variables, leads to computationally demanding iterative algorithms. Such a phenomenon becomes severe in large-scale networks having a number of antennas and users.
To handle this issue, essential requirements of practical beamforming techniques include the scalability of the network configuration, the cost-effective computation structure, and the ability to handle nonconvex objective functions.

Deep learning (DL) approaches have been recently introduced to address the challenges of existing optimization techniques \cite{Zappone:TC19, Zhang:CST19, Sun:TSP18, Xia:TC20, Kim:WCL20, Kim:CL21, Huang:TVT20, Liang:TC20, Kim:ACSSC18, Lee:JSAC19, Lee:TWC21.1}. Deep neural networks (DNNs) are employed to model an iterative process of resource management algorithm \cite{Sun:TSP18}. A supervised learning strategy is proposed that trains a DNN to learn known labels, in particular, a locally optimum power control mechanism of the WMMSE algorithm. The computational complexity is remarkably reduced compared to traditional optimization algorithms. However, since the DNN is simply trained to produce suboptimal solutions, its performance is limited by the WMMSE algorithm. 
This issue can be resolved via unsupervised learning formalisms \cite{Kim:WCL20, Liang:TC20, Kim:CL21, Huang:TVT20, Kim:ACSSC18,Lee:JSAC19, Lee:TWC21.1}. Instead of memorizing solutions of existing algorithms, these methods straightforwardly train DNNs to optimize the system performance without labels. Consequently, the unsupervised learning approach has the potential of outperforming classical optimization techniques.

These successes have triggered recent researches on the DL-based beamforming optimization \cite{Xia:TC20, Kim:WCL20, Kim:CL21, Huang:TVT20}. Convolutional neural networks (CNNs) are applied to multi-user multiple-input single-output (MU-MISO) communication systems \cite{Xia:TC20}. Channel matrices are first processed by several convolutional layers to extract the spatial correlations of fading coefficients, which is followed by fully-connected layers yielding the beamforming outputs.
The viability of the CNNs has been examined for various beamforming optimization tasks including signal-to-interference-plus-noise ratio (SINR) balancing, transmit power minimization, and sum rate maximization problems. The expert knowledge of the optimum beam structure has been exploited to design efficient DNNs for the MU-MISO systems \cite{Xia:TC20, Kim:WCL20,Kim:CL21} and MU multiple-input multiple-output (MIMO) setups \cite{Huang:TVT20}. The uplink-downlink duality formalism is included into an output activation of a final fully-connected layer. As a consequence, the DNN learns a low-dimensional representation of the optimal beamforming, and thus the architecture of the DNNs can be remarkably simplified. Such a feature learning structure has been reported to be more efficient than conventional DNN methods \cite{Kim:WCL20}. Robust beamforming optimization tasks are tackled which train DNNs with noisy channel observations \cite{Kim:CL21}. The sum rate performance is significantly improved compared to existing robust beamforming methods. Albeit their superior performance and low computational complexity, conventional DL-based beamforming solutions lack the scalability to the network size, i.e., the numbers of antennas and users. A DNN trained at a particular configuration is no longer applicable to other systems. Such an issue comes from fully-connected layers having fixed input/output dimensions. For this reason, we need to prepare a number of DNNs each dedicated to a specific combination of system parameters, which is intractable due to the prohibitive training complexity and memory requirements.

To overcome the aforementioned challenge, this paper presents a flexible learning strategy for scalable beamforming optimization problems with varying populations of antennas and users. Our goal is to identify dimensionality-invariant DNN architecture that can be universally applied to arbitrary MU-MISO systems with a single training process. The development of versatile neural layers has been recently investigated in the line of graph neural networks (GNNs) \cite{Wu:TNNLS21}. It is originally intended to handle graph-structured data samples such as citation numbers of related papers, social connections of people, and molecular structure \cite{Zhou:arXiv18}. 
Neural layers of the GNNs are carefully designed to be invariant to changes in the orders and the dimensionality of inputs and outputs. As a consequence, the GNNs can learn the interactions of distinct network entities efficiently by constructing their edge connections. Motivated by these properties, GNNs have been recently applied to the optimization of scalable wireless communication systems \cite{Shen:JSAC21,Eisen:TSP20,Lee:TWC21,Chowdhury:arXiv20,Yang:CL20,Jiang:JSAC21}. GNN-based scalable power control mechanisms have been presented in interfering networks \cite{Shen:JSAC21,Eisen:TSP20,Lee:TWC21,Chowdhury:arXiv20}. Transmitter-receiver pairs are modeled as vertices of a graph, and their interfering relationships create edges. Trainable neural parameters are shared at all vertices such that trained GNNs have good generalization ability to randomly changing network sizes, i.e., the populations of transmitter-receiver pairs. Such a parameter sharing policy leads to a flexible learning architecture that can be trained and applied to arbitrary network configurations.

There have been several efforts on GNN-based multi-antenna communication networks \cite{Shen:JSAC21,Yang:CL20,Jiang:JSAC21}. A scalable beamforming optimization task has been resolved in MISO interference channels (IFCs) \cite{Shen:JSAC21}. The weighted sum rate maximization problem is tackled with a fixed number of transmit antennas. The channel tracking problem in single-user MISO systems has been addressed using the GNN technique \cite{Yang:CL20}. Estimates of channel vectors are modeled by graph vertices, and their edges are weighted by spatial correlation factors.
The GNN is trained in a supervised manner to yield accurate estimates of actual channel vectors.
The authors in \cite{Jiang:JSAC21} consider intelligent reflecting surface (IRS) enabled MU-MISO systems. A graph consisting of an IRS and users is built for determining reflecting phases at the IRS and beamforming vectors. The scalability with respect to the user population has been examined by employing a GNN trained at a particular network setup to new configurations. Nevertheless, the adaptability to the number of antennas and reflecting elements is not viable.
These conventional GNNs are constructed on a standard graph with a homogeneous type of vertices. Therefore, their scalability is dedicated to a single category of network entities such as transmitter-receiver pairs \cite{Eisen:TSP20,Lee:TWC21} and users \cite{Jiang:JSAC21}. However, the MU-MISO system is composed of heterogeneous vertices, i.e., antennas and users, which cannot be straightforwardly tackled by existing homogeneous GNNs. To identify distinct operations for heterogeneous vertices, it is necessary to develop a novel GNN mechanism that can be applied to bipartite graphs consisting of disjoint vertex sets. There are several recent works on handling heterogeneous graph learning tasks \cite{Guo:TWC22,Zhang:arXiv21,Zhang:ACM19}. Transmit power control problems have been addressed in \cite{Guo:TWC22} for multi-cell networks where operations of antenna and user vertices are realized by the identical inference rules to determine scalar-valued optimization variables. This approach cannot be directly applied to vector-valued beamforming optimization tasks which request asymmetric antenna and user calculations. Therefore, a new GNN structure is required that is specialized to the MU-MISO systems.

Based on this intuition, we propose a bipartite graph neural network (BGNN) framework intended for multi-antenna beamforming optimization tasks. 
We first transform beamforming optimization problems into computational tasks over a bipartite graph comprising antenna vertices and user vertices. Their edge interconnections are characterized by a channel matrix where each fading coefficient acts as an attribute of an edge connecting the corresponding antenna-user pair. This motivates us to transform a scalable beamforming optimization task as a sequential decision-making process over bipartite graphs. Beamforming optimization processes are then partitioned into individual vertices whose computations are identically applied to all antennas and users. These vertex operators are designed to process scalar channel gains so that overall computations do not change with the number of antennas and users. However, such partitioned inputs and individual operations bring a suboptimal performance due to the lack of the entire channel state information. To resolve this issue, we present a bipartite message-passing (BMP) inference that facilitates pairwise interactions between antenna-user vertices through their edge connections. This allows the propagation of the partitioned vertex knowledge over the entire bipartite graph so that individual antennas and users can share useful statistics for the joint beamforming optimization. Such an iterative message-passing procedure enables the BMP inference to converge to an efficient beamforming solution only with dimensionality-invariant operations.

According to the proposed BMP inference, the optimal beamforming calculation rule can be partitioned into three different operations: message generation at antenna vertices, message generation at user vertices, and decision at user vertex. These vertex operations are implemented with three individual DNN modules, which are reused at all antenna and user vertices. This parameter sharing policy guarantees the scalability to arbitrary given antenna and user populations. These vertex DNN modules collectively form the BGNN architecture whose forwardpass computations are defined by the BMP inference. A training policy of the BGNN should be carefully designed so that the trained BGNN can be universally applied to arbitrary network configurations. To this end, the training dataset includes numerous instances of bipartite graphs, i.e., different MU-MISO configurations, each having random populations of antennas and users.
The proposed training strategy facilitates the model averaging over randomized bipartite graphs, improving the scalability and the generalization ability for unseen network setups.
The BGNN framework is applied to the sum rate and the minimum rate maximization problems. Numerical results demonstrate the scalability of the proposed BGNN method. It is shown that the knowledge of the BGNN trained at particular systems can be easily transferred to larger networks that have not been experienced in the training.

The remainder of this paper is organized as follows. Section II introduces the system model and describes scalable beamforming optimization problems. A bipartite graph-based computational inference is presented in Section III. Section IV proposes a BGNN framework which identifies scalable beamforming calculation rules. A comprehensive survey of related GNN works is presented in Section V. The effectiveness of the BGNN is demonstrated in Section VI via numerical results. The manuscript is terminated with concluding remarks in Section VII.

\textit{Notations:}
The bold capital and lowercase symbols are represented by matrix and vector, respectively. The transpose and conjugate transpose operations of a vector $\mathbf{x}$ are respectively defined as $\mathbf{x}^{T}$ and $\mathbf{x}^{H}$. We employ $\mathrm{diag}(\mathbf{x})$ to represent a diagonal matrix with elements $\mathbf{x}$. Also, $\mathbb{E}_{X}[\cdot]$ denotes an expectation operator over a random variable $X$. The sets of complex and real vectors of length $m$ are denoted by $\mathbb{C}^{m}$ and $\mathbb{R}^{m}$, respectively, and $\mathbb{C}^{m\times n}$ denotes the set of $m$-by-$n$ complex matrices.

\section{System Model} \label{sec:sec2}
We consider an MU-MISO system in which several single-antenna users are simultaneously served by a base station (BS) equipped with multiple antennas. The transmit antenna units can either be co-located at the BS or separated over a coverage area to deploy distributed antenna systems, e.g., cell-free MIMO networks \cite{Ngo:TWC17, Nayebi:TWC17}. Let $\tilde{\mathcal{N}}\triangleq\{1,\ldots,|\tilde{\mathcal{N}}|\}$ and $\tilde{\mathcal{K}}\triangleq\{1,\ldots,|\tilde{\mathcal{K}}|\}$ be the sets of all antennas and users, respectively. According to predefined antenna/user selection policies, the BS chooses an active set of antennas $\mathcal{N}\subseteq\tilde{\mathcal{N}}$ which convey wireless signals for serving a set of scheduled users $\mathcal{K}\subseteq\tilde{\mathcal{K}}$. These antenna/user selections are optimized based on the channel statistics for achieving energy-efficient transmission \cite{Khalili:TWC20} or capacity improvement \cite{Bai:TIT09}.
Therefore, the network configurations specified by the sets of active antennas $\mathcal{N}$ and users $\mathcal{K}$ should be regarded as random variables that vary at each transmission. For this reason, the beamforming strategy at the BS needs to adapt to stochastic sets $\mathcal{N}$ and $\mathcal{K}$.

Let $h_{k,i}\in\mathbb{C}^{1}$ and $v_{i,k}\in\mathbb{C}^{1}$ denote the complex channel gain and beam weight from antenna $i\in\mathcal{N}$ to user $k\in\mathcal{K}$, respectively. The received signal $y_{k}\in\mathbb{C}^{1}$ at user $k$ is written by
\begin{align}
y_{k}=\sum_{i\in\mathcal{N}}h_{k,i}v_{i,k}x_{k} + \sum_{l\in\mathcal{K}\setminus\{k\}}\left(\sum_{i\in\mathcal{N}}h_{k,i}v_{i,l}x_{l}\right) + z_{k},
\label{eq:eq1}
\end{align}
where $x_{k}\sim\mathcal{CN}(0,1)$ indicates the data symbol for user $k$ and $z_{k}\sim\mathcal{CN}(0,\sigma^{2})$ is the additive Gaussian noise at user $k$ with variance $\sigma^{2}$. Let $\mathbf{H}\in\mathbb{C}^{K\times N}\triangleq\{h_{k,i}:\forall k\in\mathcal{K},\forall i\in\mathcal{N}\}$ and $\mathbf{V}\in\mathbb{C}^{N\times K}\triangleq\{v_{i,k}:\forall i\in\mathcal{N}, \forall k\in\mathcal{K}\}$ be the collections of channel coefficients and beam weights, respectively, where $N\triangleq|\mathcal{N}|$ and $K\triangleq|\mathcal{K}|$ respectively stand for the number of active antennas and users. The achievable rate of user $k$, denoted by $r_{k}(\mathbf{H},\mathbf{V})$, is given as
\begin{align}
r_{k}(\mathbf{H},\mathbf{V})=\log_{2}\left(1+
\frac{\big|\sum_{i\in\mathcal{N}}h_{k,i}v_{i,k}\big|^{2}}{\sum_{l\in\mathcal{K}\setminus\{k\}}\big|\sum_{i\in\mathcal{N}}h_{k,i}v_{i,l}\big|^{2}+\sigma^{2}}\right).
\label{eq:eq2}
\end{align}
The network performance can be measured by the utility function $U(\cdot)$ defined over a set of the achievable rates $\{r_{k}(\mathbf{H},\mathbf{V}):\forall k\in\mathcal{K}\}$. Popular choices for $U(\cdot)$ are sum rate and minimum rate, which are respectively expressed as
\begin{align}\label{eq:eq3}
\sum_{k\in\mathcal{K}}r_{k}\left(\mathbf{H},\mathbf{V}\right)\ \text{and}\ \min_{k\in\mathcal{K}}r_{k}\left(\mathbf{H},\mathbf{V}\right).
\end{align}

\subsection{Problem Formulation} \label{subsec:subsec21}
We aim at developing a machine learning framework for scalable and universal beamforming optimization. More precisely, it is desired to optimize a beamforming operator $\mathbf{V}=\mathcal{V}(\mathbf{H})$, which maps an input channel $\mathbf{H}$ into a proper beamforming matrix $\mathbf{V}$, that can be applied to arbitrary network configurations $\mathcal{N}\subseteq\tilde{\mathcal{N}}$ and $\mathcal{K}\subseteq\tilde{\mathcal{K}}$. A generic network utility maximization problem is formulated as
\begin{subequations}\label{eq:eq4}
\begin{align}
\underset{\mathcal{V}(\cdot)}{\mathrm{max}} & \,\,\,\, \mathbb{E}_{\mathbf{H},\mathcal{N},\mathcal{K}}\left[U\left(\{r_{k}(\mathbf{H},\mathcal{V}(\mathbf{H})):\forall k\in\mathcal{K}\}\right)\right] \label{eq:eq4a}\\
\mathrm{s.t.} & \,\,\,\, \sum_{k\in\mathcal{K}}\sum_{i\in\mathcal{N}}|\mathcal{V}_{i,k}(\mathbf{H})|^{2}= P,\ \forall \mathcal{K}\subseteq\tilde{\mathcal{K}},\forall \mathcal{N}\subseteq\tilde{\mathcal{N}}, \label{eq:eq4b}
\end{align}
\end{subequations}
where $\mathcal{V}_{i,k}(\mathbf{H})\triangleq v_{i,k}$ is the $(i,k)$-th element of $\mathcal{V}(\mathbf{H})$ for a given channel $\mathbf{H}$ and $P$ in \eqref{eq:eq4b} accounts for the power constraint across all BS antennas. The objective function in \eqref{eq:eq4a} indicates the utility performance averaged over the channel distribution as well as the subsets $\mathcal{N}\subseteq\tilde{\mathcal{N}}$ and $\mathcal{K}\subseteq\tilde{\mathcal{K}}$.
Classical beamforming optimization algorithms \cite{Bjornson:SPM14, Shi:TSP11, Christensen:TWC08,Schubert:TVT04, Boche:VTC02} solve \eqref{eq:eq4} for fixed $\mathbf{H}$, $\mathcal{N}$, and $\mathcal{K}$. They require special properties for the utility function such as the convexity. Nonconvex problems would be tackled by convex approximation techniques, but these mostly bring suboptimum performance. Moreover, iterative calculation processes pose the prohibitive time complexity for large systems, thereby lacking the scalability to varying network configurations.

To address these challenges, DL-based beamforming techniques have been recently investigated in various multi-antenna systems \cite{Xia:TC20, Kim:WCL20, Kim:CL21, Huang:TVT20}. They introduce a DNN to implement the unknown beamforming operator $\mathcal{V}(\cdot)$. The DNN can be directly trained to maximize an arbitrary utility function regardless of its convexity. It has been reported that the DNN-based beamforming solution performs better than traditional optimization algorithms with much reduced complexity. However, conventional DL methods can only be applied to particular combinations of $\mathcal{N}$ and $\mathcal{K}$ that generate the training environment. These mostly rely on fully-connected layers having a fixed number of input and output variables. Therefore, DNNs trained at a certain MU-MISO system cannot be applied to other setups. For this reason, we need to prepare a number of DNNs optimized at all possible combinations of $\mathcal{N}$ and $\mathcal{K}$. However, such an approach requires prohibitive calculations in training steps as well as excessive memory requirements to store trained parameters.

\subsection{Feature Learning Structure} \label{subsec:subsec22}

We present an efficient learning strategy for the beamforming optimization. A key idea is to extract {\em beam features} that act as sufficient statistics for calculating the optimal beamforming. In particular, we desire to build a low-dimensional representation of the optimal solution to \eqref{eq:eq4} which does not incur the optimality loss. Let $\mathcal{V}_{k}(\mathbf{H})$ and $\mathbf{h}_{k}\in\mathbb{C}^{N}$ be the $k$-th column of $\mathcal{V}(\mathbf{H})$ and $\mathbf{H}^{H}$, respectively. From the dual relationship between uplink and downlink systems, the optimal beamforming vector of user $k$ is determined as \cite{Bjornson:SPM14}
\begin{align}
\mathcal{V}_{k}(\mathbf{H})=\sqrt{p_{k}}
\frac {(\sigma^{2}\mathbf{I}_{N}+\sum_{l\in\mathcal{K}}q_{l}\mathbf{h}_{l}\mathbf{h}_{l}^{H})^{-1}\mathbf{h}_{k}}{\|(\sigma^{2}\mathbf{I}_{N}+\sum_{l\in\mathcal{K}}q_{l}\mathbf{h}_{l}\mathbf{h}_{l}^{H})^{-1}\mathbf{h}_{k}\|}
\triangleq\mathcal{W}_{k}(\mathbf{H},\mathbf{s}),\label{eq:eq5}
\end{align}
where $p_{k}\geq 0$ and $q_{k}\geq 0$ respectively stand for the primal downlink power and virtual uplink power of user $k$ satisfying the power constraint $\sum_{k\in\mathcal{K}}p_{k}=\sum_{k\in\mathcal{K}}q_{k}=P$ and $\mathbf{s}\triangleq\{(p_{k},q_{k}):\forall k\in{\mathcal{K}}\}\in\mathbb{R}^{2K}$ indicates the beam feature.
It has been known that \eqref{eq:eq5} achieves all possible tradeoff points of the data rate $r_{k}(\cdot)$, $\forall k$, including the Pareto-optimal boundary \cite{Bjornson:SPM14}. Therefore, the structure \eqref{eq:eq5} becomes optimal to \eqref{eq:eq4} with arbitrary utility function $U(\cdot)$.

Let $\mathbf{s}=\mathcal{D}(\mathbf{H})$ be a computation process that yields the beam feature $\mathbf{s}$ from the channel matrix $\mathbf{H}$. Then, the beamforming operator $\mathcal{V}_{k}(\mathbf{H})$ of user $k$ can be rewritten by $\mathcal{V}_{k}(\mathbf{H})=\mathcal{W}_{k}(\mathbf{H},\mathcal{D}(\mathbf{H}))$. Substituting this into \eqref{eq:eq4} formulates an equivalent problem that finds the beam feature mapping $\mathcal{D}(\cdot)$. Without loss of the optimality, this reduces the dimension of the solution space from $2NK$
to $2K$, thereby leading to lightweight neural network architectures. For this reason, in the rest of this paper, we focus on identifying the beam feature mapping $\mathcal{D}(\cdot)$. The effectiveness of the feature learning structure \eqref{eq:eq5} has been verified in recent works \cite{Xia:TC20, Kim:WCL20, Kim:CL21, Huang:TVT20}, but its scalable learning solution has not yet been addressed. In the following sections, we propose an efficient DL framework that can solve the scalable and universal maximization problem \eqref{eq:eq4} by utilizing the beam feature $\mathbf{s}$.

\section{Bipartite Message-Passing Inference} \label{sec:sec3}

\begin{figure}
\centering
\includegraphics[width=.5\linewidth]{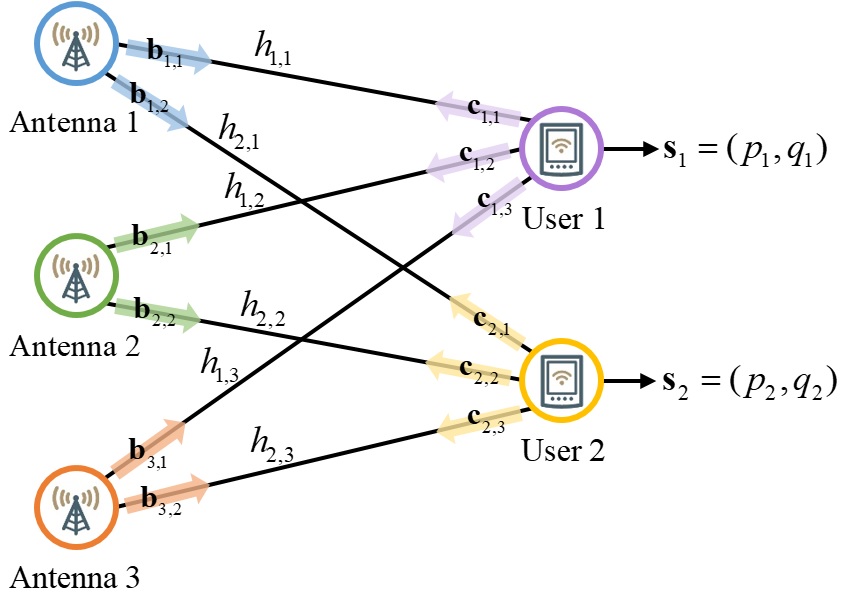}
\caption{Graphical representation of downlink multi-user networks with $N=3$ and $K=2$.}
\label{fig:fig1}
\vspace{-3mm}
\end{figure}

This section develops scalable calculation procedures of the beam feature operator $\mathbf{s}=\mathcal{D}(\mathbf{H})$ that are invariant to the dimensions of input channel matrix $\mathbf{H}\in\mathbb{C}^{K\times N}$ and output beam feature $\mathbf{s}\in\mathbb{R}^{2K}$. A key idea is to partition the operator $\mathcal{D}(\mathbf{H})$ into the smallest network entities, i.e., antennas and users, and realize their operations with individual computing modules such as neural networks.
As illustrated in Fig. \ref{fig:fig1}, we interpret the MU-MISO system as a weighted bipartite graph $\mathcal{G}$ where the antennas and users form two disjoint vertices sets. All pairs of antenna-user vertices $(i,k)$, $\forall i\in\mathcal{N},\forall k\in\mathcal{K}$, are connected through undirected edges. Due to the random fading, each antenna-user pair has different impacts on the utility function. To capture this, the relationship between antenna $i$ and user $k$ is described as a weight $h_{k,i}$ by introducing it as an attribute of edge $(i,k)$. The corresponding bipartite graph $\mathcal{G}$ can be written as a collection of disjoint vertices sets, i.e., $\mathcal{N}$ and $\mathcal{K}$, and the weighted adjacency matrix $\mathbf{H}$, i.e., $\mathcal{G}=(\mathcal{N},\mathcal{K},\mathbf{H})$.

Based on this intuition, we present a BMP inference consisting of three individual vertex operators: user message generator $\mathcal{C}(\cdot)$, antenna message generator $\mathcal{M}(\cdot)$, and feature decision maker $\mathcal{D}(\cdot)$. These vertex operators are shared at all antennas and users so that the overall computation structure becomes flexible to varying network size. Each of these component units specify calculations of antenna vertices and user vertices. In Section IV, we implement the vertex operators with neural networks. Interplays among the vertex operators lead to an iterative procedure on a bipartite graph $\mathcal{G}=(\mathcal{N},\mathcal{K},\mathbf{H})$. For scalable calculations, the vertex operators are carefully designed such that they only exploit partitioned inputs, e.g., scalar fading coefficient $h_{k,i}$, whose dimension is independent of the network size. Such a design, however, brings the lack of information for each vertex to infer the optimum beamforming computation. This requests interactions among the vertices to collect partitioned network statistics from neighbors. To this end, vertices exchange their relevant statistics, which are referred to as \textit{messages} (see Fig. \ref{fig:fig1}). Such a message-passing process is repeated several times, say $T$, so that the decision-making operator $\mathcal{D}(\cdot)$ yield a convergent beam feature $\mathbf{s}_{k}=(p_{k},q_{k})$ for each user vertex $k$. 

Let $\boldsymbol{\tau}^{[t]}$ be the quantity of a variable $\boldsymbol{\tau}$ obtained at the $t$-th iteration of the BMP inference. The derivation of the proposed BMP inference is given in Appendix \ref{appendix}. The vertex operations at the $t$-th iteration are given as follows:
\begin{subequations}\label{eq:eq6}
\begin{align}
\textit{User message generation: }&\mathbf{c}_{k,i}^{[t]} = \mathcal{C}(\mathbf{s}_{k}^{[t-1]}, \mathbf{b}_{k}^{[t-1]}, h_{k,i}), \label{eq:eq6a}\\
\textit{Antenna message generation: }&\mathbf{b}_{i,k}^{[t]}=\mathcal{M}(\mathbf{m}_{i,k}^{[t-1]}, \mathbf{c}_{i}^{[t]}, h_{k,i}), \label{eq:eq6b}\\
\textit{Beam feature decision: }&\mathbf{s}_{k}^{[t]} = \mathcal{D}(\mathbf{m}_{k}^{[t]}), \label{eq:eq6c}
\end{align}
\end{subequations}
where the vectors $\mathbf{b}_{k}^{[t]}$, $\mathbf{m}_{i,k}^{[t]}$, $\mathbf{c}_{i}^{[t]}$, and $\mathbf{m}_{k}^{[t]}$ are respectively defined as
\begin{subequations}\label{eq:eq7}
\begin{align}
\mathbf{b}_{k}^{[t]}&\triangleq\mathcal{P}(\{\mathbf{b}_{j,k}^{[t]}:\forall j\in\mathcal{N}\}),\label{eq:eq7a}\\
\mathbf{m}_{i,k}^{[t]}&\triangleq\big\{\mathbf{b}^{[t]}_{i,k},\mathcal{P}(\{\mathbf{b}_{i,l}^{[t]}:\forall l\neq k\})\big\},\label{eq:eq7b}\\
\mathbf{c}_{i}^{[t]}&\triangleq\mathcal{P}(\{\mathbf{c}_{l,i}^{[t]}:\forall l\in\mathcal{K}\}),\label{eq:eq7c}\\
\mathbf{m}_{k}^{[t]}&\triangleq\mathcal{P}(\{\mathbf{m}_{j,k}^{[t]}:\forall j\in\mathcal{N}\}), \label{eq:eq7d}
\end{align}
\end{subequations}
with $\mathcal{P}(\cdot)$ being a pooling operator. The output dimension of $\mathcal{P}(\cdot)$ is the same as that of the input. The pooling operator plays a key role in the proposed BMP inference for enabling dimensionality-invariant calculations. The detailed description of the pooling operation will be provided in Section \ref{subsec:subsec31}. 
The operators $\mathcal{C}(\cdot)$, $\mathcal{M}(\cdot)$, and $\mathcal{D}(\cdot)$ are realized by individual DNN modules. Thus, the vectors $\mathbf{c}_{k,i}^{[t]}$ in \eqref{eq:eq6a} and $\mathbf{b}_{i,k}^{[t]}$ in \eqref{eq:eq6b} act as latent vectors of each DNN module, which are optimized along with the decision variable $\mathbf{s}_{k}^{[t]}$ in \eqref{eq:eq6c}. To compute the convergent solution, these latent vectors should be sufficient statistics relevant to the final solution. In what follows, we describe the details of each step of the proposed BMP inference.

\subsection{User Message Generation} \label{subsec:subsec33}
User vertex $k$ is dedicated to the individual decision of its beam feature $\mathbf{s}_{k}=(p_{k},q_{k})$. It is obvious that for the optimal beam feature computation, the entire channel matrix $\mathbf{H}$ is required at all user vertices. However, such an input lacks the scalability on the network size $N$ and $K$. Thus, the user operator $\mathcal{C}(\cdot)$ in \eqref{eq:eq6a} is designed to get the scalar channel gain $h_{k,i}$. The resultant output is the user message vector denoted by $\mathbf{c}_{k,i}^{[t]}\in\mathbb{R}^{M}$, where the hyperparameter $M$ stands for the dimension of the message. The vector $\mathbf{c}_{k,i}^{[t]}$ acts as a representation of the knowledge at user $k$ that is requested from antenna $i$. The user message $\mathbf{c}_{k,i}^{[t]}$ is forwarded to the associated antenna vertex $i\in\mathcal{N}$. To calculate this, we input all the available information to the operator $\mathcal{C}(\cdot)$. At the $t$-th iteration, user $k$ has its previous decision $\mathbf{s}_{k}^{[t-1]}\triangleq(p_{k}^{[t-1]},q_{k}^{[t-1]})$. Also, the vector $\mathbf{b}_{k}^{[t-1]}\in\mathbb{R}^{M}$ in \eqref{eq:eq7a} is assumed to be available which encodes a set of past messages $\{\mathbf{b}_{j,k}^{[t-1]}:\forall j\in\mathcal{N}\}$ obtained from antenna vertices using the pooling operator $\mathcal{P}(\cdot)$. Aggregating all these variables, user $k$ creates the message vectors $\mathbf{c}_{k,i}^{[t]}$, $\forall i\in\mathcal{N}$, for all interconnected antennas by using the operator $\mathcal{C}(\cdot)$ in \eqref{eq:eq6a}. Consequently, the operator $\mathcal{C}(\cdot)$ maps an input data of length $M+4$ to the $M$-dimensional output message vector, which is no longer dependent on $N$ and $K$.

\subsection{Antenna Message Generation} \label{subsec:subsec32}
The antennas act as intermediate relay nodes that fuse and broadcast the partitioned information of the users. Thus, it allows the indirect message exchange between user vertices that are not straightforwardly connected in the bipartite graph. To this end, antenna $i$ collects the user message $\mathbf{c}_{k,i}^{[t]}$ from all connected user vertices $k\in\mathcal{K}$. These are leveraged to create the intermediate information vector $\mathbf{b}^{[t]}_{i,k}\in\mathbb{R}^{M}$ using the antenna operator $\mathcal{M}(\cdot)$ in \eqref{eq:eq6b}, which characterizes the knowledge about user $k$ inferred at antenna $i$. We encapsulate the associated channel gain $h_{k,i}$ as well as the past information vectors $\mathbf{m}_{i,k}^{[t-1]}\in\mathbb{R}^{2M}$ and $\mathbf{c}_{i}^{[t]}\in\mathbb{R}^{M}$ in \eqref{eq:eq7c}. As a consequence, the input dimension of the antenna operator $\mathcal{M}(\cdot)$ is given by $3M+2$, which is the constant with respect to $N$ and $K$. The final antenna message $\mathbf{m}_{i,k}^{[t]}$, which will be conveyed to user $k\in\mathcal{K}$, is constructed according to \eqref{eq:eq7b}. This contains all the information about user $k$ including the dedicated vector $\mathbf{b}^{[t]}_{i,k}$ and the leakages $\{\mathbf{b}_{i,l}^{[t]}:\forall l\neq k\}$ intended for other users.

\subsection{Beam Feature Decision} \label{subsec:subsec31}
The decision for the beam feature $\mathbf{s}^{[t]}_{k}$ is taken at individual user $k$. The decision operator $\mathcal{D}(\cdot)$ in \eqref{eq:eq6c} leverages the aggregation $\mathbf{m}_{k}^{[t]}\in\mathbb{R}^{2M}$ in \eqref{eq:eq7d} of all antenna messages $\{\mathbf{m}_{i,k}^{[t]}:\forall i\in\mathcal{N}\}$.
A series of vertex operations \eqref{eq:eq6a}, \eqref{eq:eq6b}, and \eqref{eq:eq6c} for $t=1,\ldots,T$ leads to the final decision $\mathbf{s}_{k}^{[T]}$. By collecting all features $\{\mathbf{s}_{k}^{[T]}:\forall k\in\mathcal{K}\}$, we can obtain the beamforming vectors $\mathbf{V}$ from \eqref{eq:eq5}. We summarize the proposed BMP inference in Algorithm \ref{Algorithm}.

\begin{algorithm}[htp]
\caption{BMP inference}\label{Algorithm}
\begin{algorithmic}
\STATE Initialize $\mathbf{s}_{k}^{[0]}$, $\mathbf{b}_{i,k}^{[0]}$, $\forall i \in\mathcal{N}$, $\forall k \in\mathcal{K}$, and $t=0$.
\FOR{$t=1,2,\ldots,T$}
\STATE \textit{1. User message generation: }
\STATE \hskip1.2em\indent User $k$ creates a message $\mathbf{c}_{k,i}^{[t]}$ from \eqref{eq:eq6a} and sends it to antenna $i\in\mathcal{N}$.
\STATE \textit{2. Antenna message generation: }
\STATE \hskip1.2em\indent Antenna $i$ creates a message $\mathbf{b}_{i,k}^{[t]}$ from \eqref{eq:eq6b} and sends it to user $k\in\mathcal{K}$.
\STATE \textit{3. Beam feature decision: }
\STATE \hskip1.2em\indent User $k$ takes a decision $\mathbf{s}_{k}^{[t]}$ from \eqref{eq:eq6c}.
\ENDFOR
\STATE Calculate the beamforming vectors $\mathcal{V}(\mathbf{H})$ from \eqref{eq:eq5}.
\end{algorithmic}
\end{algorithm}

The proposed BMP inference relies on the pooling operator $\mathcal{P}(\cdot)$ in \eqref{eq:eq7} which generates an aggregation of the received messages. This needs to be carefully designed such that it ensures the dimensionality-invariant calculations of the vertex operators $\mathcal{C}(\cdot)$, $\mathcal{M}(\cdot)$, and $\mathcal{D}(\cdot)$. To this end, its computation should be defined on the set of messages, e.g., $\{\mathbf{b}_{j,k}^{[t]}:\forall j\in\mathcal{N}\}$ in \eqref{eq:eq7a}, but not on their simple concatenation. Such a property makes $\mathcal{P}(\cdot)$ become the set function that is independent of the antenna set $\mathcal{N}$ and user set $\mathcal{K}$, more precisely, the bipartite graphs comprising them. There have been various choices for such a pooling operator. In this work, we adopt the sum pooling operator which has been known to be efficient in various applications \cite{Shen:JSAC21, Lee:TWC21, Yang:CL20, Chowdhury:arXiv20}. Let $\tau_{u}$ be a variable of vertex $u\in\mathcal{U}$. For some mappings $\rho(\cdot)$ and $\phi(\cdot)$, the following structure can characterize a continuous set function $\mathcal{P}(\{\tau_{u}:\forall u\in\mathcal{U}\})$ \cite{Lee:TWC21,MZaheer:17}.
\begin{align}
    \mathcal{P}(\{\tau_{u}:\forall u\in\mathcal{U}\})=\rho\left(\sum_{u\in\mathcal{U}}\phi(\tau_{u})\right).\label{eq:pooling}
\end{align}
It has been revealed that the dimensionality constraint on the inner mapping $\phi:\mathbb{R}\rightarrow\mathbb{R}^{|\mathcal{U}|+1}$ builds the universal approximation property. More precisely, it verifies the existence of the mappings $\rho(\cdot)$ and $\phi(\cdot)$ that accurately model the unknown optimum set function. Thus, we can use the structure in \eqref{eq:pooling} to realize the pooling operator $\mathcal{P}(\cdot)$ of the proposed BMP inference. By substituting \eqref{eq:pooling} into \eqref{eq:eq6}, the BMP inference now consists of the inner mapping $\phi(\cdot)$ and outer mapping $\rho(\cdot)$. For instance, \eqref{eq:eq6a} is modified as
\begin{align}\label{eq:eq6ex}
\mathbf{c}_{k,i}^{[t]} = \mathcal{C}\left(\mathbf{s}_{k}^{[t-1]}, \rho\left(\sum_{j\in\mathcal{N}}\phi(\mathbf{b}_{j,k}^{[t-1]})\right), h_{k,i}\right).
\end{align}
Notice that $\rho(\cdot)$ and $\phi(\cdot)$ can be abstracted into the preceding and succeeding operators $\mathcal{C}(\cdot)$, $\mathcal{M}(\cdot)$, and $\mathcal{D}(\cdot)$, respectively. Since these operators are characterized by DNNs, we do not need to design $\rho(\cdot)$ and $\phi(\cdot)$ explicitly. Therefore, the pooling operator $\mathcal{P}(\{\tau_{u}:\forall u\in\mathcal{U}\})$ in \eqref{eq:eq6ex} can be simplified as the summation $\mathcal{P}(\{\tau_{u}:\forall u\in\mathcal{U}\})=\sum_{u\in\mathcal{U}}\tau_{u}$. As a result, the BMP inference can be universally applied to arbitrary bipartite graphs $\mathcal{G}=(\mathcal{N},\mathcal{K},\mathbf{H})$.

\section{Bipartite Graph Neural Network} \label{sec:sec4}
This section presents a BGNN framework that realizes the BMP inference in Algorithm \ref{Algorithm} with the aid of DNNs.
We model the vertex operators in \eqref{eq:eq6} as parameterized intelligences, in particular, fully-connected DNNs. For an input vector $\mathbf{u}\in\mathbb{R}^{N_{0}}$ of length $N_{0}$, computations of an $L$-layer DNN $\mathcal{F}_{L}(\cdot;\boldsymbol{\Theta})$ with a trainable parameter set $\boldsymbol{\Theta}$ are given as
\begin{align} \label{eq:eq8}
\mathcal{F}_{L}\left(\mathbf{u};\boldsymbol{\Theta}\right)=\delta_{L}\left(\mathbf{W}_{L}\times\cdots\times\delta_{1}\left(\mathbf{W}_{1}\mathbf{u}+\mathbf{o}_{1}\right)+\cdots+\mathbf{o}_{L}\right),
\end{align}
where $\delta_{l}(\cdot)$ is the activation function of layer $l$ ($l=1,\ldots,L$), $\mathbf{W}_{l}\in\mathbb{R}^{N_{l-1}\times N_{l}}$ and $\mathbf{o}_{l}\in\mathbb{R}^{N_{l}}$ respectively represent the weight matrix and bias vector of layer $l$, which collectively form the trainable parameter set $\boldsymbol{\Theta}\triangleq\{\mathbf{W}_{l}, \mathbf{o}_{l}:\forall l\}$, and $N_{l}$ indicates the output dimension of layer $l$.
The operators $\mathcal{C}(\cdot)$, $\mathcal{M}(\cdot)$, and $\mathcal{D}(\cdot)$ are respectively designed by DNNs $\mathcal{F}_{L_{\mathcal{C}}}(\cdot ; \boldsymbol{\Theta}_{\mathcal{C}})$, $\mathcal{F}_{L_{\mathcal{M}}}(\cdot ; \boldsymbol{\Theta}_{\mathcal{M}})$, and $\mathcal{F}_{L_{\mathcal{D}}}(\cdot; \boldsymbol{\Theta}_{\mathcal{D}})$ as
\begin{subequations}\label{eq:eq9}
\begin{align}
&\mathbf{c}_{k,i}^{[t]}=\mathcal{C}(\mathbf{s}_{k}^{[t-1]},\mathbf{b}_{k}^{[t-1]},h_{k,i})=\mathcal{F}_{L_{\mathcal{C}}}(\mathbf{s}_{k}^{[t-1]},\mathbf{b}_{k}^{[t-1]},h_{k,i} ;\boldsymbol{\Theta}_{\mathcal{C}}), \label{eq:eq9a} \\
&\mathbf{b}_{i,k}^{[t]}=\mathcal{M}(\mathbf{m}_{i,k}^{[t-1]},\mathbf{c}_{i}^{[t]},h_{k,i})=\mathcal{F}_{L_{\mathcal{M}}}(\mathbf{m}_{i,k}^{[t-1]},\mathbf{c}_{i}^{[t]},h_{k,i} ;\boldsymbol{\Theta}_{\mathcal{M}}), \label{eq:eq9b} \\
&\mathbf{s}_{k}^{[t]}=\mathcal{D}(\mathbf{m}_{k}^{[t]})=\mathcal{F}_{L_{\mathcal{D}}}(\mathbf{m}_{k}^{[t]} ;{\boldsymbol{\Theta}}_{\mathcal{D}}). \label{eq:eq9c}
\end{align}
\end{subequations}
The DNNs in \eqref{eq:eq9} parameterize unstructured computational inferences $\mathcal{C}(\cdot)$, $\mathcal{M}(\cdot)$, and $\mathcal{D}(\cdot)$. The accuracy of such approximations can be verified by the universal approximation theorem \cite{Hornik:NN89}. It states that a properly designed DNN can approximate any continuous-valued function within an arbitrary small approximation error. This is valid for unknown optimal mappings $\mathcal{C}^{\star}(\cdot)$, $\mathcal{M}^{\star}(\cdot)$, and $\mathcal{D}^{\star}(\cdot)$ of problem \eqref{eq:eq4}, thereby ensuring the optimality of our DNN models \eqref{eq:eq9}. 

Combining \eqref{eq:eq9} results in a trainable beamforming optimizer $\mathbf{V}^{[t]}(\boldsymbol{\Theta})$ with parameter $\boldsymbol{\Theta}\triangleq\{\boldsymbol{\Theta}_{\mathcal{C}}, \boldsymbol{\Theta}_{\mathcal{M}},\boldsymbol{\Theta}_{\mathcal{D}}\}$, which is defined as
\begin{align}
\mathbf{V}^{[t]}(\boldsymbol{\Theta})
\triangleq\mathcal{W}(\mathbf{H},\mathbf{s}^{[t]}), \label{eq:eq10}
\end{align}
where $\mathcal{W}(\mathbf{H},\mathbf{s})\triangleq\{\mathcal{W}_{k}(\mathbf{H},\mathbf{s}):\forall k\in\mathcal{K}\}$ denotes the collection of the beamforming structure in \eqref{eq:eq5}. Among a sequence of the beam features $\mathbf{s}^{[1]},\ldots,\mathbf{s}^{[T]}$, we choose the last value $\mathbf{s}^{[T]}$ for obtaining the final beamforming solution $\mathbf{V}^{[T]}$. Plugging \eqref{eq:eq10} into \eqref{eq:eq4} leads to the training problem expressed as
\begin{align}\label{eq:eq11}
\underset{\boldsymbol{\Theta}}{\mathrm{max}}\ \ \mathcal{U}^{[T]}(\boldsymbol{\Theta})\triangleq \mathbb{E}_{\mathbf{H},\mathcal{N},\mathcal{K}}\left[U\left(\{r_{k}(\mathbf{H},\mathbf{V}^{[T]}(\boldsymbol{\Theta})):\forall k\in\mathcal{K}\}\right)\right],
\end{align}
where $\mathcal{U}^{[t]}(\boldsymbol{\Theta})$ stands for the average utility achieved by the BGNN output $\mathbf{s}^{[t]}$ of the $t$-th iteration. The optimization variable now turns into the set of trainable parameters $\boldsymbol{\Theta}$ with closed-form forwardpass computations specified by Algorithm \ref{Algorithm} and DNNs \eqref{eq:eq9}. It is thus more tractable than the original formulation \eqref{eq:eq4} that invokes the identification process of unstructured mapping~$\mathcal{V}(\cdot)$.

\subsection{Training Policy} \label{subsec:subsec41}

The optimization of the BGNN parameter $\boldsymbol{\Theta}$ is achieved via existing training algorithms, e.g., the mini-batch stochastic gradient descent (SGD) methods.
The forwardpass of the BGNN is constructed by the BMP inference in Algorithm 1 which involves iterative operations with $T$ repetitions. Such a sequential computation is desired to learn the convergent message-passing rules so that a sequence of the beam features $\mathbf{s}^{[1]},\ldots,\mathbf{s}^{[T]}$ approaches the optimum point. At the same time, the BMP inference integrates three DNNs in \eqref{eq:eq9} into a deep architecture having $T$ layers. Increasing $T$ enhances the expressive power of the BGNN, which is, in general, beneficial to find complicated beamforming optimization rules. However, it is not trivial to train a very deep structure due to the gradient vanishing issue where backpropagation algorithm fails to get valid gradients of the BGNN parameter $\boldsymbol{\Theta}$ and gets stuck into a poor stationary point.

The gradient vanishing issue can be resolved by involving skip connections between hidden layers, e.g., residual modules of the ResNet \cite{He:CVPR16}. To inject this idea, we propose a new training objective function that includes utility values $\mathcal{U}^{[t]}(\boldsymbol{\Theta})$ at all iteration steps $t=1,\ldots,T$. The proposed training objective function $F(\boldsymbol{\Theta})$ is given by
\begin{align}\label{eq:eq12}
F(\boldsymbol{\Theta})=\sum_{t=1}^{T}\mathcal{U}^{[t]}(\boldsymbol{\Theta})=\sum_{t=1}^{T}\mathbb{E}_{\mathbf{H},\mathcal{N},\mathcal{K}}\left[U\left(\{r_{k}(\mathbf{H},\mathbf{V}^{[t]}(\boldsymbol{\Theta})):\forall k\in\mathcal{K}\}\right)\right].
\end{align}
The objective function in \eqref{eq:eq12} captures a complete trajectory of all beam features $\mathbf{s}^{[t]}$, $\forall t$, instead of focusing only on the final decision $\mathbf{s}^{[T]}$ as in \eqref{eq:eq11}. By doing so, we can provide various shortcuts to the error backpropagation algorithm, thereby transferring valid gradients in the backwardpass. In the proposed BGNN, the vertex DNNs \eqref{eq:eq9} are designed to accept their previous outputs as the additional input data, which form the concatenation-based skip connections suggested in the DenseNet \cite{Huang:CVPR17}. The number of skip connections gets larger as the number of antennas $N$ and users $K$ grow. This facilitates an efficient training of the BGNN and its scalability for larger networks. In addition, the proposed training objective adjusts the intermediate beam feature $\mathbf{s}^{[t]}$ jointly along with its past and next decisions for maximizing the aggregated utility values $\sum_{t=1}^{T}\mathcal{U}^{[t]}(\boldsymbol{\Theta})$.
Such an incremental decision process helps the BGNN gradually improve the utility performance $\mathcal{U}^{[t]}(\boldsymbol{\Theta})$ as the iteration step $t$ grows.

The maximization problem of the training objective \eqref{eq:eq12} can be solved by using the mini-batch SGD algorithm and its variants, e.g., the Adam algorithm \cite{Kingma:ICLR15}. At each training epoch, the parameter set $\boldsymbol{\Theta}$ is updated as
\begin{align}
\boldsymbol{\Theta}\leftarrow\boldsymbol{\Theta} + \eta\mathbb{E}_{\mathcal{B}}\left[\sum_{t=1}^{T}\nabla_{\boldsymbol{\Theta}}U\left(\{r_{k}(\mathbf{H},\mathbf{V}^{[t]}(\boldsymbol{\Theta})):\forall k\in\mathcal{K}\}\right)\right], \label{eq:eq13}
\end{align}
where $\eta>0$ indicates the learning rate and $\nabla_{\tau}$ stands for the gradient operator with respect to a variable $\tau$. Here, $\mathcal{B}$ is the mini-batch set containing numerous instances of the weighted bipartite graphs $\mathcal{G}=(\mathcal{N},\mathcal{K},\mathbf{H})$ that are collected in advance, or, drawn from known distributions.

\begin{algorithm}[htp]
\caption{BGNN Training Algorithm}\label{Algorithm2}
\begin{algorithmic}
\STATE Initialize $\boldsymbol{\Theta}$ and $\mathcal{U}_{\text{Best}}=\infty$.
\REPEAT
\STATE 1. Generate the mini-batch set $\mathcal{B}$ containing random bipartite graphs $\mathcal{G}=(\mathcal{N},\mathcal{K},\mathbf{H})$.
\STATE 2. Update the parameter set $\boldsymbol{\Theta}$ from \eqref{eq:eq13}.
\STATE 3. Obtain the validation performance $\mathcal{U}_{\text{Val}}$ using independent validation samples.
\IF{$\mathcal{U}_{\text{Val}}\geq\mathcal{U}_{\text{Best}}$ }
\STATE Save the BGNN parameter $\boldsymbol{\Theta}$.
\ENDIF
\UNTIL{convergence}
\end{algorithmic}
\end{algorithm}

Algorithm \ref{Algorithm2} summarizes the training policy of the proposed BGNN. At each training epoch, we first generate the mini-batch set $\mathcal{B}$ containing random bipartite graphs $\mathcal{G}=(\mathcal{N},\mathcal{K},\mathbf{H})$. In particular, the sets of antennas $\mathcal{N}$ and users $\mathcal{K}$ are uniformly chosen from the entire sets $\tilde{\mathcal{N}}$ and $\tilde{\mathcal{K}}$, respectively. Then, for given cardinalities $N=|\mathcal{N}|$ and $K=|\mathcal{K}|$, the corresponding channel samples $\mathbf{H}\in\mathbb{C}^{K\times N}$ are randomly generated from, e.g., the Rayleigh distribution. Training the BGNN with random bipartite graphs can be viewed as the dropout technique \cite{Srivastava:14} where hidden neurons of a DNN are randomly removed during the training. The proposed training strategy stochastically activates the vertex DNN and their edge connections according to the training samples $\mathcal{G}$. By doing so, we perform the model averaging over a number of MU-MISO configurations, which improves the generalization ability of the trained BGNN to unseen network setups.
The sample mean of the gradient in \eqref{eq:eq13} is computed via the backpropagation algorithm. After the update of the BGNN parameter $\boldsymbol{\Theta}$, we evaluate the validation objective value, denoted by $\mathcal{U}_{\text{Val}}$, over independently generated validation samples. Then, we store the parameter set whenever a better validation performance is obtained. The proposed training strategy can be categorized into the unsupervised learning since it does not rely on labels, i.e., the optimum solution of \eqref{eq:eq4}. Therefore, the DNNs in \eqref{eq:eq9} can learn appropriate beamforming rules by themselves, instead of memorizing known suboptimal points as in the supervised learning method \cite{Sun:TSP18}. The training process is executed in an offline manner. Online calculations are realized by the trained DNN modules by means of linear matrix multiplications \eqref{eq:eq8}.

\subsection{Decentralized and Parallel Implementation} \label{subsec:subsec43}
We discuss a decentralized implementation strategy of the trained BGNN. This is particularly beneficial for distributed MIMO configurations, e.g., cell-free MIMO and fog radio access networks, where transmit antennas are separately deployed. Furthermore, it facilitates parallel operations across individual antennas and users, thereby leading to a computationally-efficient implementation. The component DNNs in \eqref{eq:eq9} are installed at their desired nodes by storing the trained parameters. Since the identical DNNs are applied to all antennas and users, we simply add or remove the sets of the DNN modules whenever the network configurations change. For real-time beamforming optimization, the antennas and users participate in a cooperative decision process using the BMP inference in Algorithm \ref{Algorithm}. The channel coefficients $\{h_{k,i}:\forall k\in\mathcal{K},\forall i\in\mathcal{N}\}$ are first acquired via standard channel estimation procedures. With the aid of the channel reciprocity or feedback mechanism, $h_{k,i}$ can be obtained at the associated antenna $i$ and user $k$. User $k$ generates the message $\mathbf{c}_{k,i}^{[t]}$ and sends it to the intended antenna $i$ through reliable control channels. Receiving all uplink messages $\mathbf{c}_{k,i}^{[t]}$, $\forall k\in\mathcal{K}$, antenna $i$ creates its messages $\mathbf{b}_{i,k}^{[t]}$, $\forall k\in\mathcal{K}$, and forwards them to the associated users. These messages can be utilized as control signals exchanged among antennas and users for calculating the beam features recursively. Only the final decision $\mathbf{s}_{k}^{[T]}$ is fed to the BS for retrieving the beamforming vectors $\mathbf{V}$. The beamforming computation tasks are distributed to individual antennas and users in a parallel manner. As a consequence, the online computational complexity of the trained BGNN is generally lower than that of the existing centralized learning solutions and classical optimization techniques. This will be clearly shown in the simulation section.

\subsection{Beam Recovery for Minimum Rate Utility} \label{subsec:subsec42}
In general, the beam recovery process in \eqref{eq:eq5} requires both the downlink and uplink power variables $\mathbf{p}$ and $\mathbf{q}$. In the case of the minimum rate maximization problem, we can optimally retrieve the beamforming vectors only using the uplink power vector $\mathbf{q}$ \cite{Schubert:TVT04}. Thus, the beam feature can be set to the scalar variable, i.e., $s_{k}=q_{k}$. This halves the output dimension of the BGNN compared to the structure in \eqref{eq:eq5}, thereby leading to a more efficient learning strategy.

Based on this intuition, we propose a new feature learning strategy intended for the minimum rate utility $\min_{k\in\mathcal{K}}r_{k}(\mathbf{H},\mathbf{V})$. The DNN $\mathcal{F}_{L_{\mathcal{D}}}(\cdot;\boldsymbol{\Theta}_{\mathcal{D}})$ in \eqref{eq:eq9c} outputs the uplink power $q_{k}$ only. These collectively form the beam direction $\tilde{\mathbf{v}}_{k}\in\mathbb{C}^{N}$ of user $k$ expressed~as
\begin{align}\label{eq:sinr}
\tilde{\mathbf{v}}_{k}=\frac{(\sigma^{2}\mathbf{I}_{N}+\sum_{l\in\mathcal{K}}q_{l}\mathbf{h}_{l}\mathbf{h}_{l}^{H})^{-1}\mathbf{h}_{k}}{\|(\sigma^{2}\mathbf{I}_{N}+\sum_{l\in\mathcal{K}}q_{l}\mathbf{h}_{l}\mathbf{h}_{l}^{H})^{-1}\mathbf{h}_{k}\|}.
\end{align}
We then calculate two diagonal matrices $\boldsymbol{\Omega}\in\mathbb{R}^{K\times K}$ and $\mathbf{\Phi}\in\mathbb{R}^{K\times K}$. First, $\boldsymbol{\Omega}$ is given as $\boldsymbol{\Omega}=\mathrm{diag}(\gamma_{1}/|\mathbf{h}_{1}^{H}\tilde{\mathbf{v}}_{1}|^{2}, \ldots, \gamma_{K}/|\mathbf{h}_{K}^{H}\tilde{\mathbf{v}}_{K}|^{2})$ with $\gamma_{k}$ being the target SINR value of user $k$. Also, the matrix $\boldsymbol{\Phi}$ contains all the zero value of the diagonal terms and off-diagonal term is consisted of $|\mathbf{h}_{k}^{H}\tilde{\mathbf{v}}_{l}|^{2}$, $k\neq l$. 
Using these, we obtain the matrix $\boldsymbol{\Gamma}\in\mathbb{R}^{(K+1)\times(K+1)}$ given~as
\begin{align}\label{eq:eq14}
\boldsymbol{\Gamma} = \left[\begin{array}{cc}\boldsymbol{\Omega}\boldsymbol{\Phi} & \sigma^{2}\boldsymbol{\Omega}\boldsymbol{1}_{K}\\\frac{1}{P}\mathbf{1}_{K}^{T}\boldsymbol{\Omega}\boldsymbol{\Phi} & \frac{\sigma^{2}}{P}\text{tr}(\boldsymbol{\Omega})\end{array}\right]
\end{align}
with $\mathbf{1}_{K}$ being the all-one column vector of length $K$. According to \cite{Schubert:TVT04}, the downlink power vector $\mathbf{p}$ becomes the solution of the following eigenvalue decomposition (EVD) problem:
\begin{align}\label{eq:EVD}
\boldsymbol{\Gamma}
\begin{bmatrix}
\mathbf{p} \\
1
\end{bmatrix}
=\lambda_{\max}
\begin{bmatrix}
\mathbf{p} \\
1
\end{bmatrix}
\end{align}
where $\lambda_{\max}$ is the maximum eigenvalue of the matrix $\boldsymbol{\Gamma}$. By using \eqref{eq:sinr}-\eqref{eq:EVD}, we can optimally recover the downlink power $\mathbf{p}$ from the dual uplink power $\mathbf{q}$. The final beamforming vector $\mathbf{v}_{k}$ is then obtained as $\mathbf{v}_{k}=\sqrt{p_{k}}\tilde{\mathbf{v}}_{k}$.

A similar feature learning structure has been recently presented in \cite{Xia:TC20}. This work pursues a supervised learning policy where a DNN simply memorizes known labels, i.e., the optimal uplink power $\mathbf{q}$ generated by classical optimization algorithms. Thus, the nonlinear operations in \eqref{eq:sinr}-\eqref{eq:EVD} are not included in training and designing of the DNN. In contrast, the proposed framework addresses unsupervised learning problems so that the BGNN pioneers an efficient optimization policy of the unknown optimal uplink power. To inject this concept, the BGNN should be trained along with \eqref{eq:sinr}-\eqref{eq:EVD}. Their gradients are available in closed-form expressions, and thus we can train the BGNN through existing backpropagation algorithm. However, the computation complexity of the gradient of the EVD \eqref{eq:EVD} has the cubic order with respect to the user population $K$. 
We address this challenge by using a new training objective function. It is inferred from \cite{Bjornson:SPM14} that the downlink SINR values can be equivalently represented by the SINR of the dual uplink system. This allows us to transform the utility function \eqref{eq:eq4a} of the original downlink system into the dual utility function defined over dual uplink rate values. As a consequence, the training objective evaluated in the dual uplink system becomes a function of the uplink transmit power $\mathbf{q}$ only, but not of the downlink one $\mathbf{p}$. This constructs a new backpropagation path which straightforwardly calculates the gradient of the BGNN without the EVD. 
We will verify the advantage of the learning structure in Section \ref{subsec:subsec52}.

\section{Related Works} \label{subsec:subsec44}
According to the parameter sharing strategy, the GNN can be categorized into two different classes \cite{Wu:TNNLS21}: graph recurrent neural networks (GRNNs) \cite{Yang:CL20,Lee:TWC21} and graph convolutional neural networks (GCNNs) \cite{Eisen:TSP20,Shen:JSAC21,Chowdhury:arXiv20,Jiang:JSAC21}. The proposed BGNN falls into the GRNN framework which reuse the identical DNN parameters both in the spatial and temporal domains. 
On the contrary, the GCNN leverages distinct DNNs at individual iteration steps.  
The GRNN is effective to solve sequential graph representation tasks \cite{Ruiz:TSP20}. The BGNN also learns a sequential decision-making process of the optimum beam feature over bipartite graphs. This leads to the BMP inference in Algorithm \ref{Algorithm} that exploits the identical computational operators at all iteration steps. Such a strategy has been widely adopted in classical beamforming optimization algorithms, e.g., the WMMSE method, which adopts the same recursive update rules. The GRNN policy removes unnecessary trainable parameters, thereby providing efficient DNN construction.

Both the GRNN and GCNN approaches have been widely adopted in addressing the scalability issue of large-scale wireless networks.
The GRNN in \cite{Lee:TWC21} handles power control problems in single antenna IFCs by characterizing transmitter-receiver pairs as vertices.
Various network management problems have been tackled based on the GCNN \cite{Eisen:TSP20}. This method models arbitrary interfering wireless networks as random edge graphs whose vertices correspond to transmitter-receiver pairs. 
The work in \cite{Chowdhury:arXiv20} adopts the unfolding technique to derive model-based learning strategy that mimics existing power control algorithms. Each layer of the unfolded structure is implemented by the GCNN. The message passing GNN (MPGNN) \cite{Shen:JSAC21} extends the GCNN to the beamforming optimization problems in the MISO IFC. Similar to \cite{Lee:TWC21,Chowdhury:arXiv20,Eisen:TSP20}, transmitter-receiver pairs are realized with vertices. The MPGNN effectively scales up with the link populations, but it is applicable only with the fixed number of antennas.

The aforementioned GNNs focus on a learning task over homogeneous graphs consisting of a single type of network entities, e.g., transmitter-receiver pairs \cite{Lee:TWC21,Chowdhury:arXiv20,Eisen:TSP20}, receive antennas \cite{Yang:CL20}, and users \cite{Jiang:JSAC21}. On the contrary, we need to consider bipartite graphs having two different types of vertices, i.e., antenna and user vertices. 
Such a challenge has been recently addressed by the heterogeneous GNN (HetGNN) structure \cite{Zhang:arXiv21,Guo:TWC22,Zhang:ACM19}. This method leverages different DNNs for each of different vertex types. By doing so, we can exploit an additional design degree-of-freedom that characterizes heterogeneous operations of different types of vertices. The HetGNN has been applied to network management problems \cite{Zhang:arXiv21,Guo:TWC22}. 
The beamforming optimization scheme is proposed in \cite{Zhang:arXiv21} for the MISO IFC. According to the number of antennas, this framework considers two different types of transmitters, i.e., single-antenna and two-antenna transmitters. However, likewise the conventional homogeneous GNNs \cite{Lee:TWC21,Chowdhury:arXiv20,Eisen:TSP20}, vertices of each type model transmitter-receiver pairs. For this reason, it still lacks the flexibility to the antenna populations.

This issue has been resolved by using the HetGNN proposed in \cite{Guo:TWC22} which aims at solving the power control problem in general multi-cell MU-MISO systems. The BSs and users are respectively characterized as different types of vertices in bipartite graphs. 
The message-generating inferences $\mathcal{M}(\cdot)$ and $\mathcal{C}(\cdot)$ and decision-making inference $\mathcal{D}(\cdot)$ of \cite{Guo:TWC22} are given as
\begin{subequations} \label{eq:eqHetGNN1}
\begin{align} 
&\mathbf{b}_{i,k}^{[t]} = \mathcal{M}(\mathbf{b}_{i,k}^{[t]}, \mathbf{c}_{i}^{[t-1]})= \delta\left(\mathbf{S}\mathbf{b}_{i,k}^{[t-1]}+\mathbf{c}_{i}^{[t-1]}\right),
\label{eq:eqHetGNN1a} \\
&\mathbf{c}_{k,i}^{[t]} = \mathcal{C}(\mathbf{c}_{k,i}^{[t]}, \mathbf{b}_{k}^{[t-1]})= \delta\left(\mathbf{T}\mathbf{c}_{k,i}^{[t-1]}+\mathbf{b}_{k}^{[t-1]}\right), \label{eq:eqHetGNN1b} \\
&v_{i,k} = \mathcal{D}(\mathbf{b}_{i,k}^{[T]}), \label{eq:eqHetGNN1c}
\end{align}
\end{subequations}
where $\delta(\cdot)$ represents the activation and $\mathbf{b}_{k}^{[t]}$ and $\mathbf{c}_{i}^{[t]}$ are respectively obtained as
\begin{subequations} 
\begin{align} 
\mathbf{b}_{k}^{[t]}\triangleq\sum\nolimits_{j=1}^{N}\sum\nolimits_{l=1}^{K}(\mathbf{C}\mathbf{c}_{l,j}^{[t-1]}+\mathbf{P}h_{l,i})\ \text{and}\ \mathbf{c}_{i}^{[t]}\triangleq\sum\nolimits_{l=1}^{K}\sum\nolimits_{j=1}^{N}(\mathbf{D}\mathbf{b}_{j,l}^{[t-1]}+\mathbf{Q}h_{k,j}), \nonumber
\end{align}
\end{subequations}
with $\{\mathbf{S},\mathbf{C},\mathbf{P},\mathbf{T},\mathbf{D},\mathbf{Q}\}$ being the set of trainable parameters. Here, the scalar beam weight $v_{i,k}$ is retrieved as in \eqref{eq:eqHetGNN1c} only at the final iteration step $T$. The proposed BMP inference in \eqref{eq:eq6} has several distinct features compared with \eqref{eq:eqHetGNN1}. First, \cite{Guo:TWC22} applies similar computation structures for the message generation rules at antenna vertex \eqref{eq:eqHetGNN1a} and user vertex \eqref{eq:eqHetGNN1b}. 
This strategy would fail to capture heterogeneous roles of antennas and users.
On the contrary, the proposed BGNN adopts dedicated inferences to different types of vertices that are specially designed for the beamforming optimization problems. Consequently, the proposed learning structure is more suitable to the MU-MISO systems. Second, in the BGNN, the user vertices are responsible for calculating the beam feature $\mathbf{s}_{k}^{[t]}$, whereas the HetGNN recovers the beam weights $v_{i,k}$ directly at the antenna vertex. Thus, to obtain the entire beam weights $\{v_{i,k}:\forall k\in\mathcal{K}\}$, each antenna should repeat the decision inference \eqref{eq:eqHetGNN1c} $K$ times.
In contrast, the proposed BGNN simply takes a single decision variable $\mathbf{s}_{k}^{[t]}$ only using a sole DNN unit. This leads to a lightweight implementation of the learning structure.

The notion of graph filters (GFs) \cite{Eisen:TSP20,Segarra:TSP17,Gama:SPM20,Isufi:TPAMI} can also be employed for bipartite graph learning tasks. This framework builds explainable GNN models whose effectiveness is rigorously proved through the graph signal processing technique. 
The feasibility of the node-invariant GFs has been recently proved in various networking applications \cite{Eisen:TSP20}. 
It realizes multi-hop graph signal processing by applying scalar GFs so that all vertices employ the identical filter coefficients. 
The node-variant GF \cite{Segarra:TSP17} adopts different GFs at each of vertices, thereby allowing heterogeneous vertex operations. Still, however, both the node-invariant and node-variant GFs fail to model the asymmetry of neighboring vertices.
This issue can be resolved by using edge-varying GFs \cite{Gama:SPM20,Isufi:TPAMI} where each vertex assigns distinct GFs to different neighbors. The importance of individual neighbors can be asymmetrically reflected in calculating vertex messages via weighted aggregation operations. This feature is indeed captured by the proposed BGNN which employs different set of DNNs at antenna vertices and user vertices. Extending the GF-based approaches to the MU-MISO systems would be an interesting future work.

\section{Numerical Results} \label{sec:sec5}
This section evaluates the performance of the proposed BGNN framework for the sum rate and minimum rate utilities in \eqref{eq:eq3}.
Users are randomly dropped in a circle cell of radius $100$ m, and the BS antennas are co-located at the center of the cell. The channel coefficient $h_{k,i}$ is the complex Gaussian random variable with zero mean and variance $\rho_{k}$, which indicates the large-scale fading of user $k$. We set $\rho_{k}=1/(1+(d_{k}/d_{\text{ref}})^{\alpha})$ \cite{Kim:WCL20, Kim:CL21} where $d_{k}$ is the distance between the BS and user $k$, $d_{\text{ref}}=30$ m and $\alpha=3$ stand for the reference distance and the path-loss exponent, respectively. For the unit noise variance $\sigma^{2}=1$, the SNR equals $P$.

\begin{table}[ht]
\centering
\caption{DNN setup}\label{tab:table1}
\subtable[$\mathcal{F}_{L_{\mathcal{C}}}(\cdot;\boldsymbol{\Theta}_{\mathcal{C}})$ and $\mathcal{F}_{L_{\mathcal{M}}}(\cdot;\boldsymbol{\Theta}_{\mathcal{M}})$]{
\begin{tabular}{lccll}
\hline\hline
             & \begin{tabular}[c]{@{}c@{}}Output dimension\\ \text{[Sum rate]}\end{tabular} & \multicolumn{3}{c}{\begin{tabular}[c]{@{}c@{}}Output dimension\\ \text{[Minimum rate]}\end{tabular}} \\ \hline
Input        & 200                                                                   & \multicolumn{3}{c}{40}                                                                        \\
Dense+ReLU & 200                                                                   & \multicolumn{3}{c}{40}                                                                        \\
Dense+Tanh & $M$                                                                   & \multicolumn{3}{c}{$M$}                                                                       \\ \hline\hline
\end{tabular}
}
\subtable[$\mathcal{F}_{L_{\mathcal{D}}}(\cdot;\boldsymbol{\Theta}_{\mathcal{D}})$]{
\begin{tabular}{lccll}
\hline\hline
             & \begin{tabular}[c]{@{}c@{}}Output dimension\\ \text{[Sum rate]}\end{tabular} & \multicolumn{3}{c}{\begin{tabular}[c]{@{}c@{}}Output dimension\\ \text{[Minimum rate]}\end{tabular}} \\ \hline
Input        & 200                                                                   & \multicolumn{3}{c}{40}                                                                        \\
Dense+ReLU & 200                                                                   & \multicolumn{3}{c}{40}                                                                        \\
Dense+Sigmoid & 2                                                                   & \multicolumn{3}{c}{1}                                                                       \\ \hline\hline
\end{tabular}
}
\end{table}

The architectures of the DNNs $\mathcal{F}_{L_{\mathcal{C}}}(\cdot;\boldsymbol{\Theta}_{\mathcal{C}})$, $\mathcal{F}_{L_{\mathcal{M}}}(\cdot;\boldsymbol{\Theta}_{\mathcal{M}})$, and $\mathcal{F}_{L_{\mathcal{D}}}(\cdot;\boldsymbol{\Theta}_{\mathcal{D}})$ are given in Table~\ref{tab:table1}. The activation functions at all hidden layers are fixed as the rectified linear unit (ReLU). Output layers of the DNNs $\mathcal{F}_{L_{\mathcal{C}}}(\cdot;\boldsymbol{\Theta}_{\mathcal{C}})$ and $\mathcal{F}_{L_{\mathcal{M}}}(\cdot;\boldsymbol{\Theta}_{\mathcal{M}})$ are set to the hyperbolic tangent function, whereas that of $\mathcal{F}_{L_{\mathcal{D}}}(\cdot;\boldsymbol{\Theta}_{\mathcal{D}})$ is fixed as the sigmoid function.
The message dimension $M$, i.e., the length of the message vectors $\mathbf{b}_{i,k}^{[t]}$ and $\mathbf{c}_{k,i}^{[t]}$, is given as $M=5$. Elements of initial messages are drawn from the independent Gaussian distribution with zero mean and unit variance. The Adam algorithm \cite{Kingma:ICLR15} is adopted with the learning rate $\eta=0.0005$ and the mini-batch size $\mathcal{B}=1000$. In the training, the maximum populations of the antennas and users are set to $|\tilde{\mathcal{N}}|=|\tilde{\mathcal{K}}|=8$. 
The BGNN is trained over 100 epochs where each epoch consists of $50$ mini-batch sets.
Validation and test procedures are carried out with $5\times10^{3}$ independent samples. The simulations are implemented with Tensorflow 1.15.0 on a PC with Intel i7-7700K CPU, 64 GB RAM, and a Titan XP GPU.

In what follows, we focus on validating the proposed BGNN framework in the co-located antenna case where all transmit antenna ports are installed at the BS. Two popular utility functions are examined: the sum rate utility in Section \ref{subsec:subsec51} and the minimum rate utility function in Section \ref{subsec:subsec52}. Then, it is extended to address the beamforming optimization task in cell-free MIMO systems in Section \ref{subsec:subsec53} in which antenna ports are separately deployed over the cell.

\subsection{Sum Rate Utility} \label{subsec:subsec51}

\begin{figure}
\centering
    \subfigure[$P=10$ dB, $N=4$]{
        \includegraphics[width=.33\linewidth]{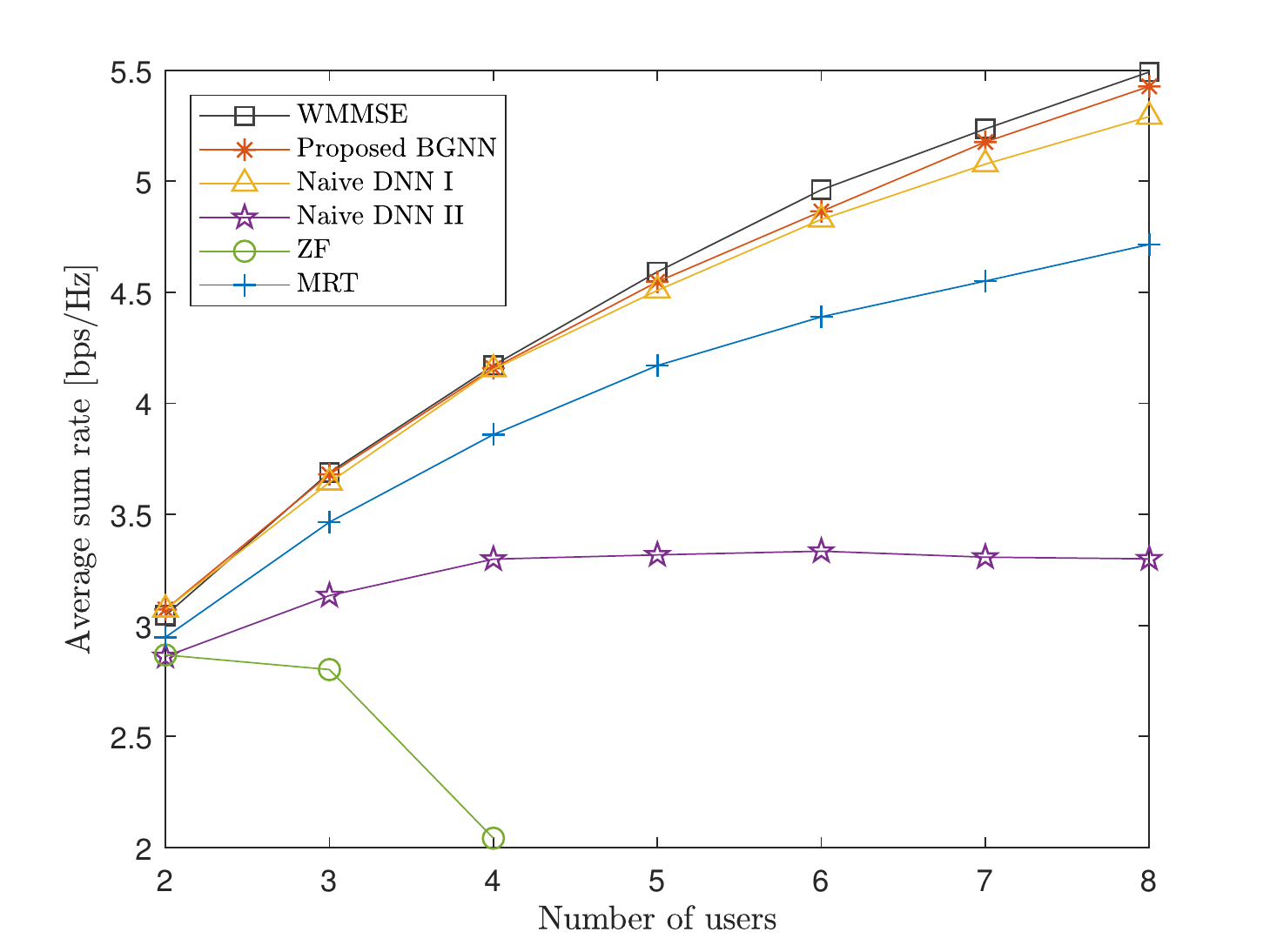}\label{fig:fig2a}
    }\hspace{-5mm}
    \subfigure[$P=10$ dB, $N=6$]{
        \includegraphics[width=.33\linewidth]{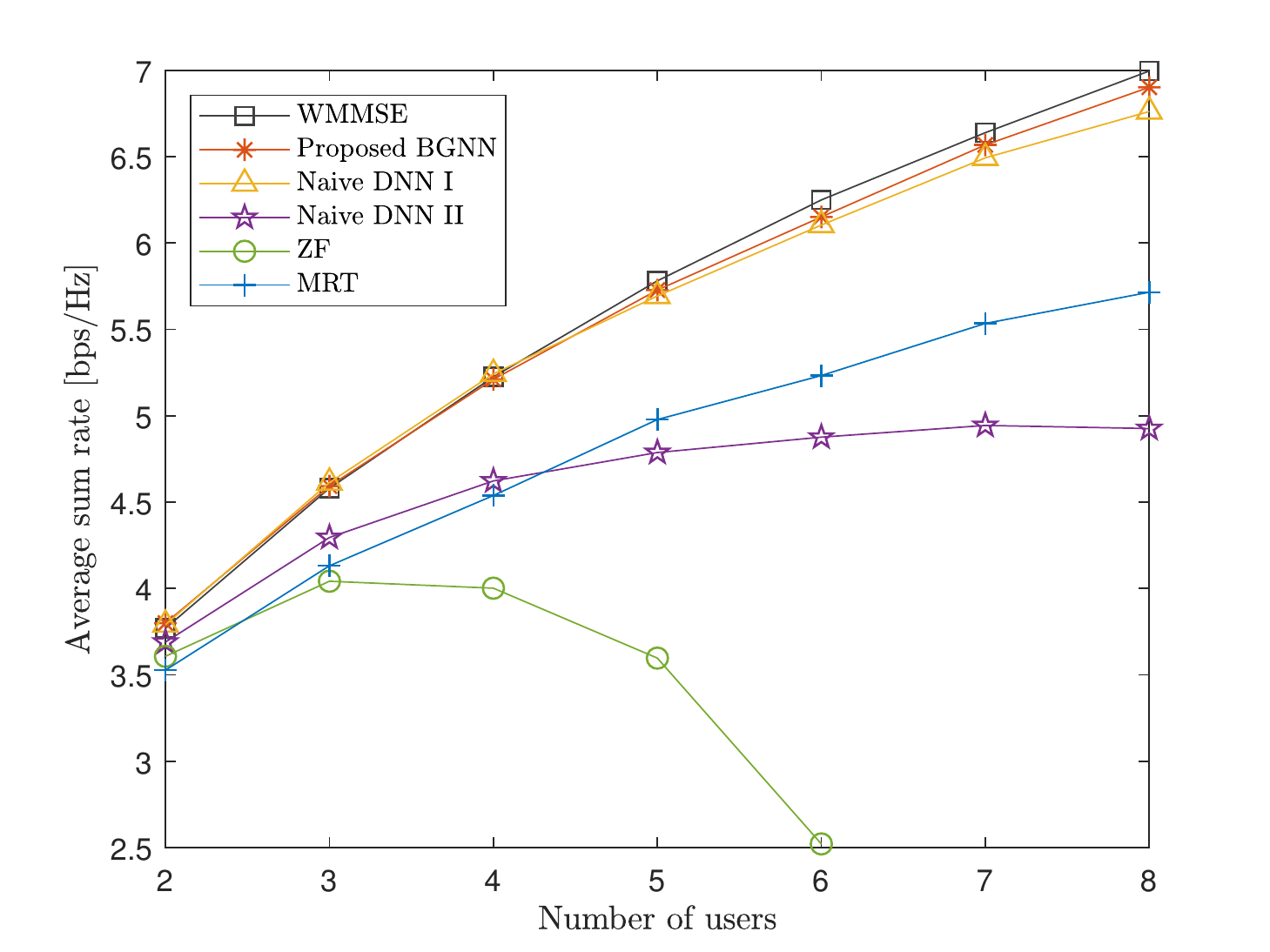}\label{fig:fig2b}
    }\hspace{-5mm}
    \subfigure[$P=10$ dB, $N=8$]{
        \includegraphics[width=.33\linewidth]{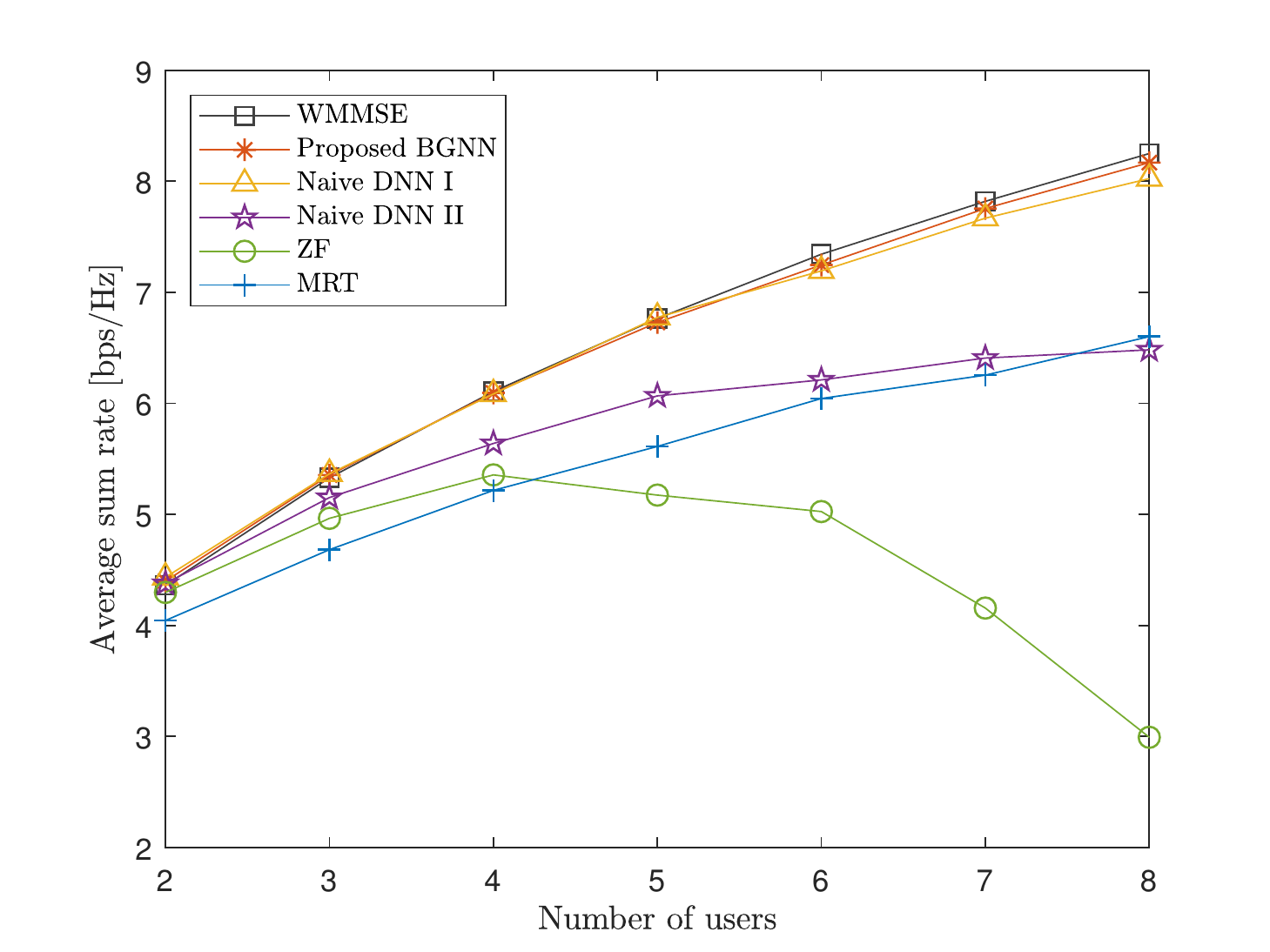}\label{fig:fig2c}
    }
    \subfigure[$P=25$ dB, $N=4$]{
        \includegraphics[width=.33\linewidth]{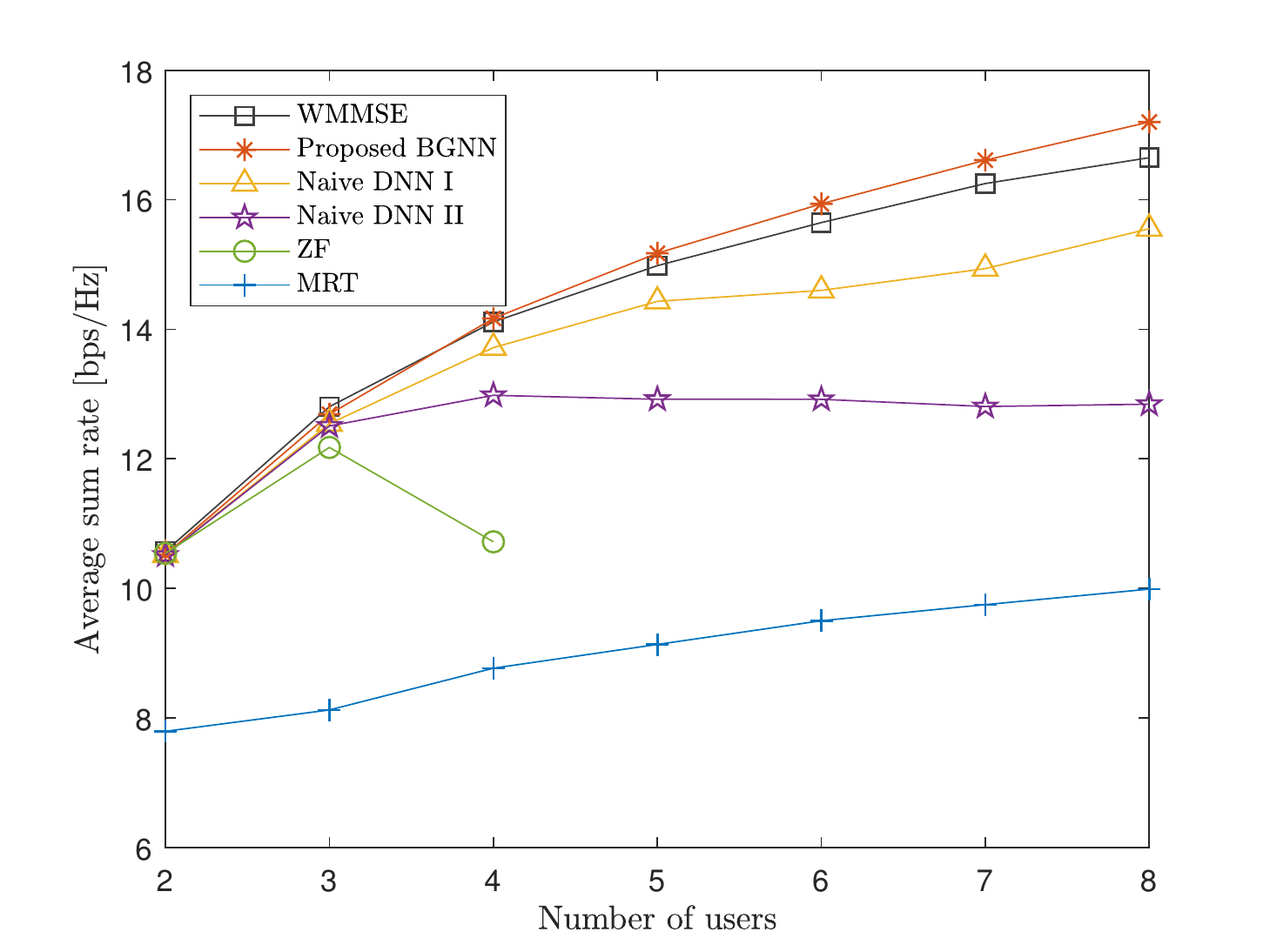}\label{fig:fig2d}
    }\hspace{-5mm}
    \subfigure[$P=25$ dB, $N=6$]{
        \includegraphics[width=.33\linewidth]{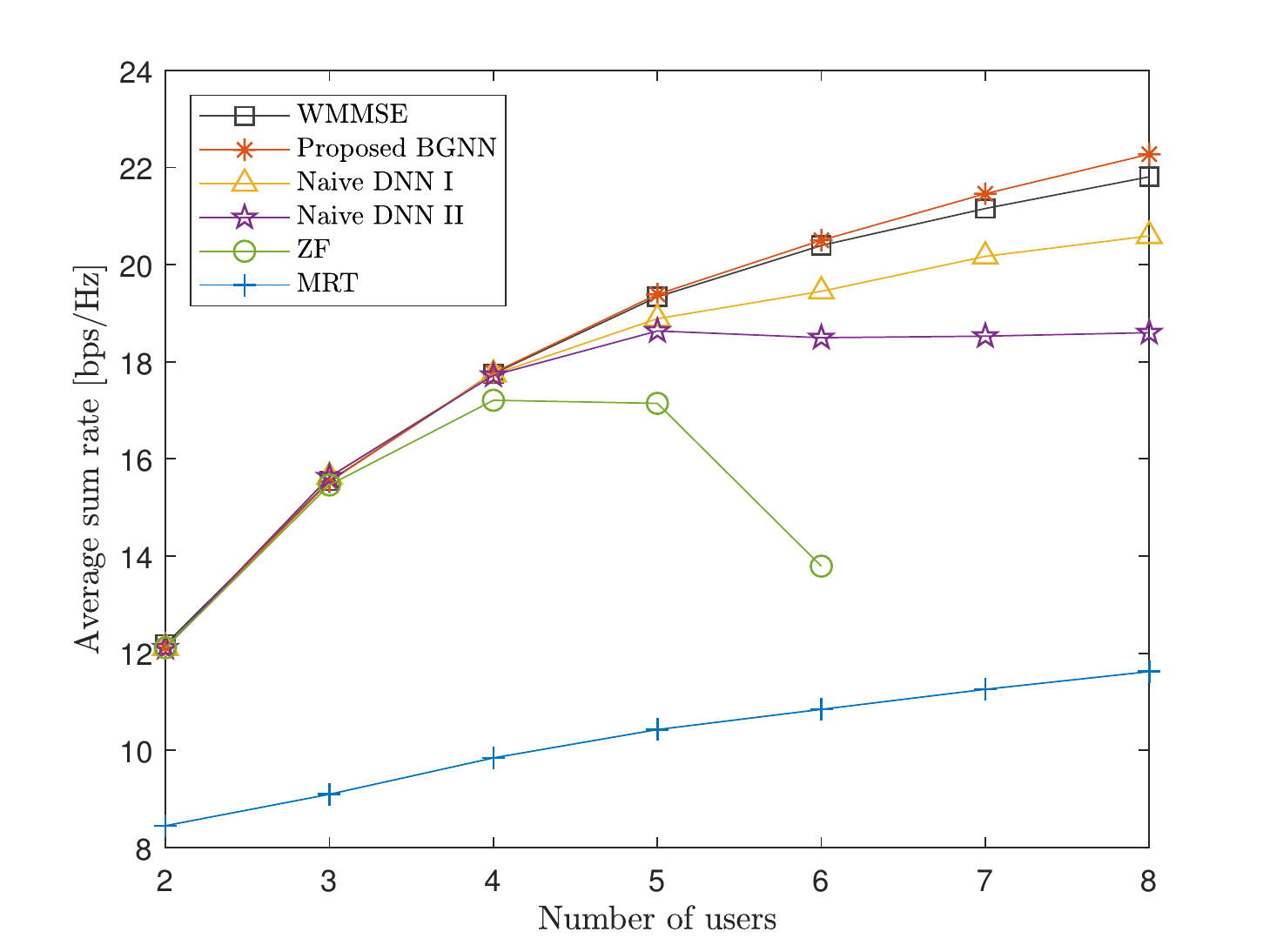}\label{fig:fig2e}
    }\hspace{-5mm}
    \subfigure[$P=25$ dB, $N=8$]{
        \includegraphics[width=.33\linewidth]{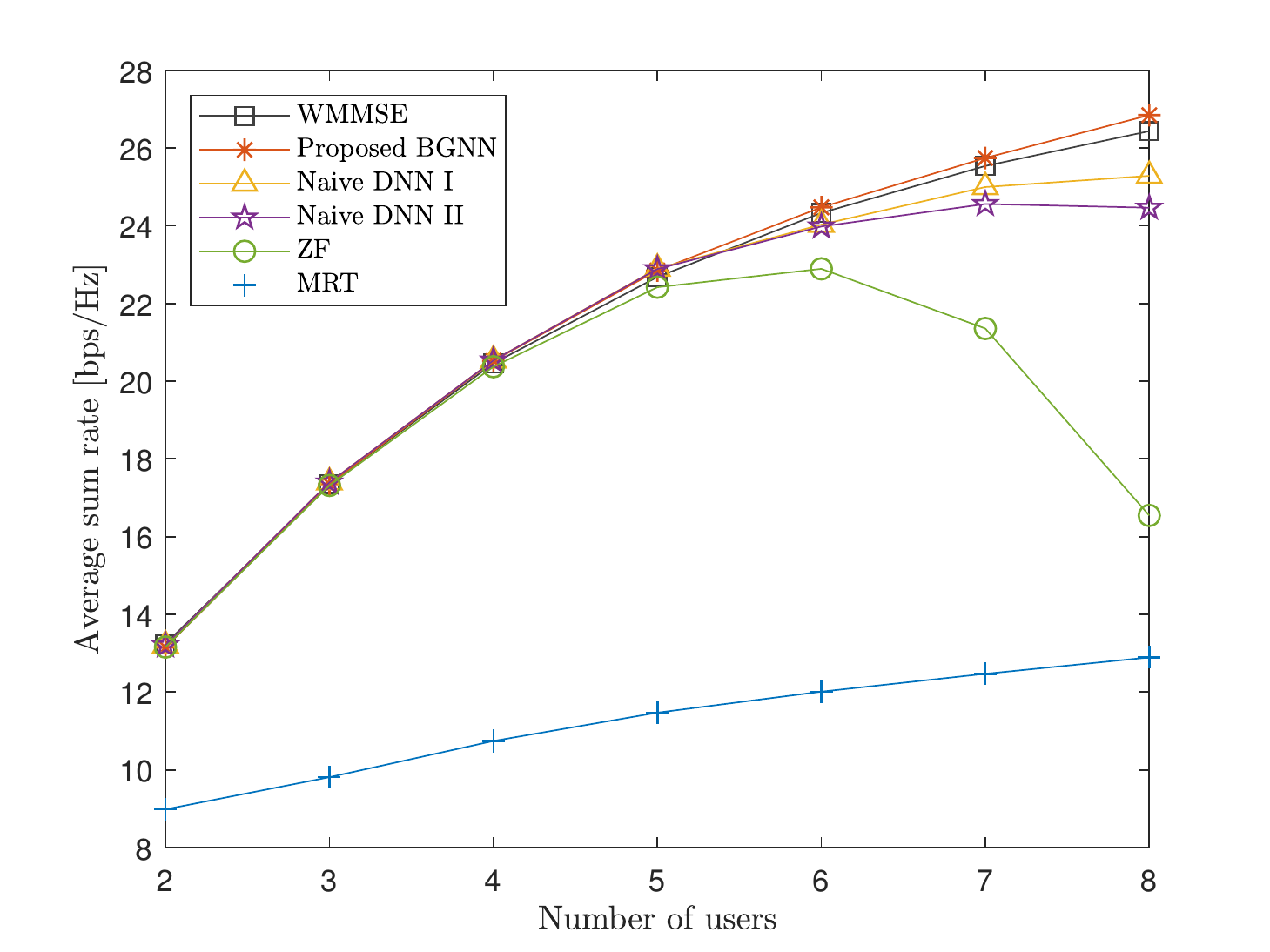}\label{fig:fig2f}
    }
    \caption{Average sum rate performance in various network configurations.}
    \label{fig:fig2}
\end{figure}

The following benchmark schemes are considered.
\begin{itemize}
\item \textit{WMMSE: }Locally optimum beamforming vectors are obtained by using the WMMSE algorithm \cite{Christensen:TWC08} for each realization of test bipartite graph samples $(\mathcal{N},\mathcal{K},\mathbf{H})$.
\item \textit{ZF/MRT: }Closed-form zero-forcing (ZF) and maximum ratio transmission (MRT) beamforming vectors are utilized with the optimized transmit power allocation.
\item \textit{Naive DNN:} Conventional fully-connected DNN processes the entire channel matrix $\mathbf{H}$ directly \cite{Kim:WCL20}. Due to the fixed computational structure, multiple DNNs should be trained for all combinations of $\mathcal{N}$ and $\mathcal{K}$.
\end{itemize}
Two different types of the naive DNN baselines are taken into account. First, ``Naive DNN I'' employs the identical number of trainable parameters to the proposed BGNN, thereby resulting in a similar memory utilization. On the contrary, ``Naive DNN II'' method has the same level of the depth and the width of the BGNN so that they show a similar level of the expressive power.

Fig. \ref{fig:fig2} demonstrates the scalability of the proposed BGNN by evaluating the test sum rate performance for various network configurations. We depict the average sum rate versus the number of users $K$ for $N\in\{4,6,8\}$. It is noted that the ZF baseline becomes infeasible when $N<K$. First, we can see that the BGNN exhibits superior performance to the benchmark schemes for all simulated configurations. It performs better than the locally optimum WMMSE algorithm in the high SNR regime ($P=25$ dB). The naive DNN baselines should conduct on multiple training tasks each dedicated to a particular combination of $N$ and $K$, thereby leading to prohibitive training complexity and memory requirement. Nevertheless, their sum rate performance is degraded compared to the proposed scheme especially in large systems. The performance gap between the BGNN and ``Naive DNN I'' schemes increases as the number of users $K$ gets larger although they require identical memory storage. In addition, a very deep structure of the ``Naive DNN II'' method fails to learn in all simulated configurations and becomes even worse than a shallow DNN of the ``Naive DNN I'' method. This proves the effectiveness of the proposed learning policy over simple fully-connected structures.

\begin{figure}
\centering
    \subfigure[$P=10$ dB]{
        \includegraphics[width=.35\linewidth]{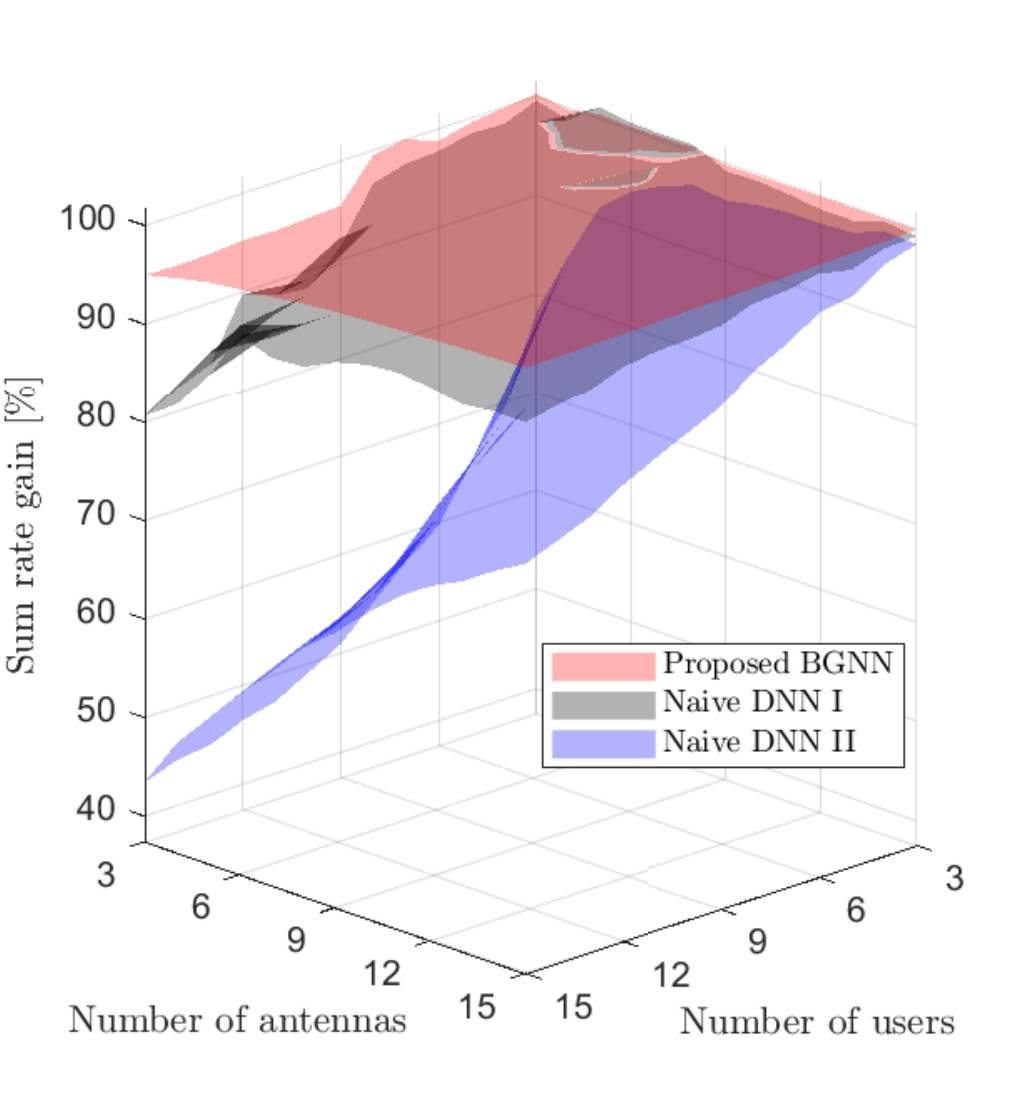}\label{fig:fig3a}
    }\hspace{5mm}
    \subfigure[$P=25$ dB]{
        \includegraphics[width=.35\linewidth]{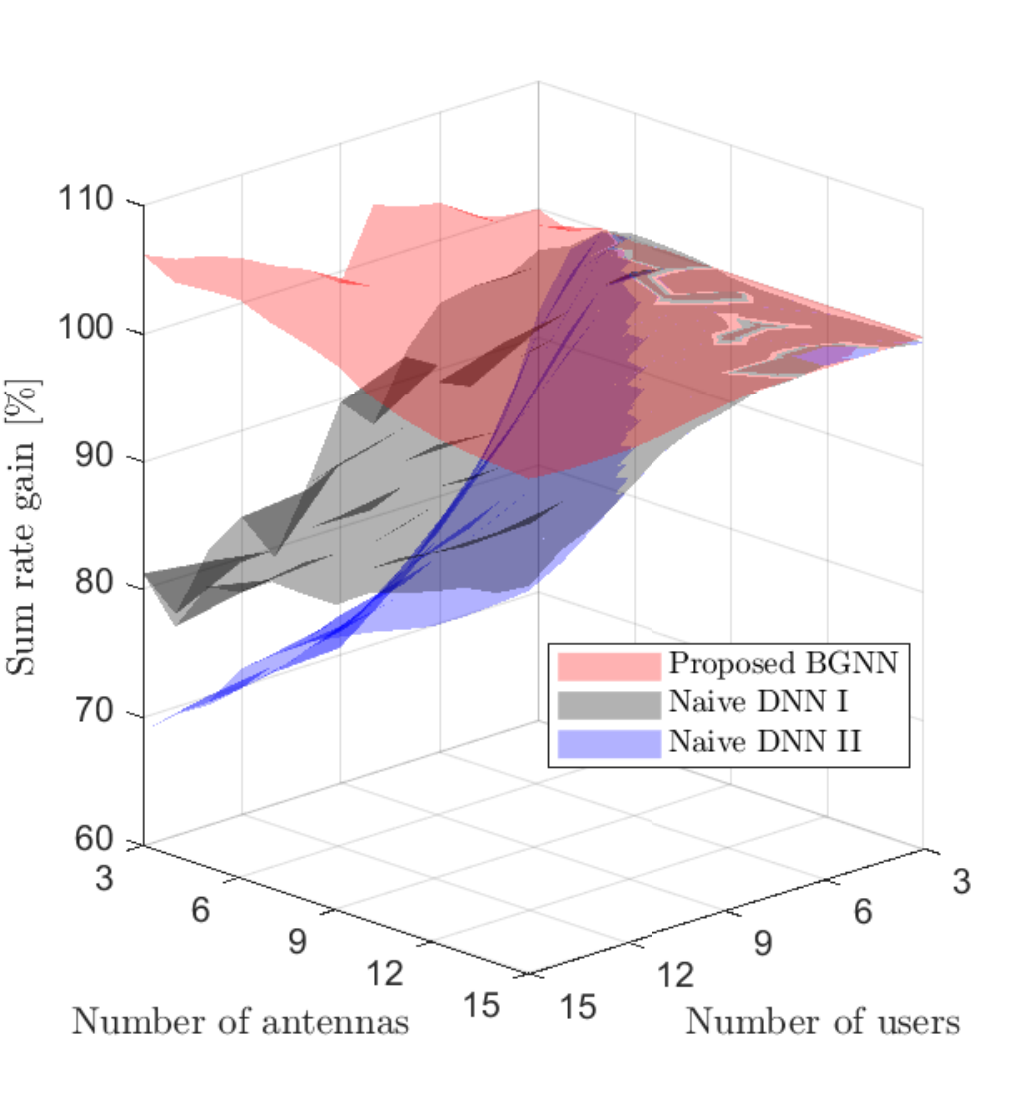}\label{fig:fig3b}
    }
    \caption{Average sum rate gain with respect to $N$ and $K$.}
    \label{fig:fig3}
\end{figure}

We further examine the scalability of the BGNN in Fig. \ref{fig:fig3} which shows the average sum rate gain  by varying $N$ and $K$ for $P\in\{10, 25\}$ dB. Here, the sum rate gain is defined as the sum rate performance normalized by that of the WMMSE method. The BGNN trained with the maximum antenna/user populations $N=K=8$ is straightforwardly applied to larger systems up to $N=K=15$ antennas and users. Thus, total $133$ configurations for $N,K\in[9,15]$ are not available in the training step. Nevertheless, the proposed BGNN provides almost identical performance to the WMMSE algorithm in these unseen network setups. On the contrary, the conventional DL baselines, which are trained at dedicated system configurations, exhibit significant performance losses. This validates the generalization ability of the BGNN in the sense that it indeed pioneers efficient and universal beamforming optimization rules, not just simply memorizing input-output relationships for specific systems as in conventional works.

\begin{figure}
\centering
\includegraphics[width=.45\linewidth]{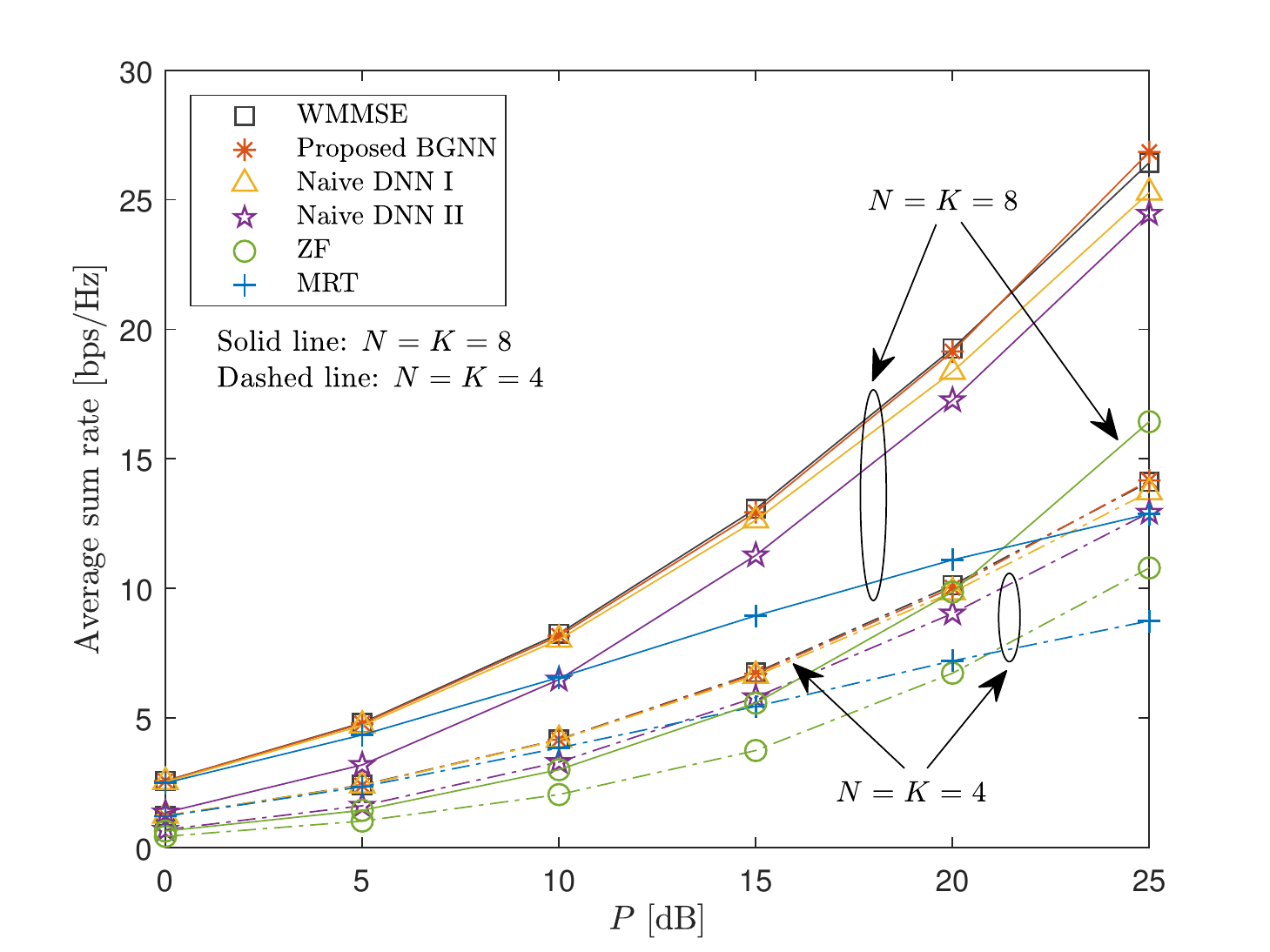}
\caption{The average sum rate versus the SNR $P$ with $N=K=4$ and $8$.}
\label{fig:fig4}
\end{figure}

Fig. \ref{fig:fig4} presents the average sum rate performance with respect to the SNR $P$ for $N=K=4$ and $8$. For all simulated SNR regime, the BGNN is superior to the DNN-based baseline schemes and achieves almost identical performance of the locally optimum WMMSE algorithm. Thus, we can conclude that the proposed approach performs well in a wide range of SNRs.

\begin{figure}
\centering
\includegraphics[width=.45\linewidth]{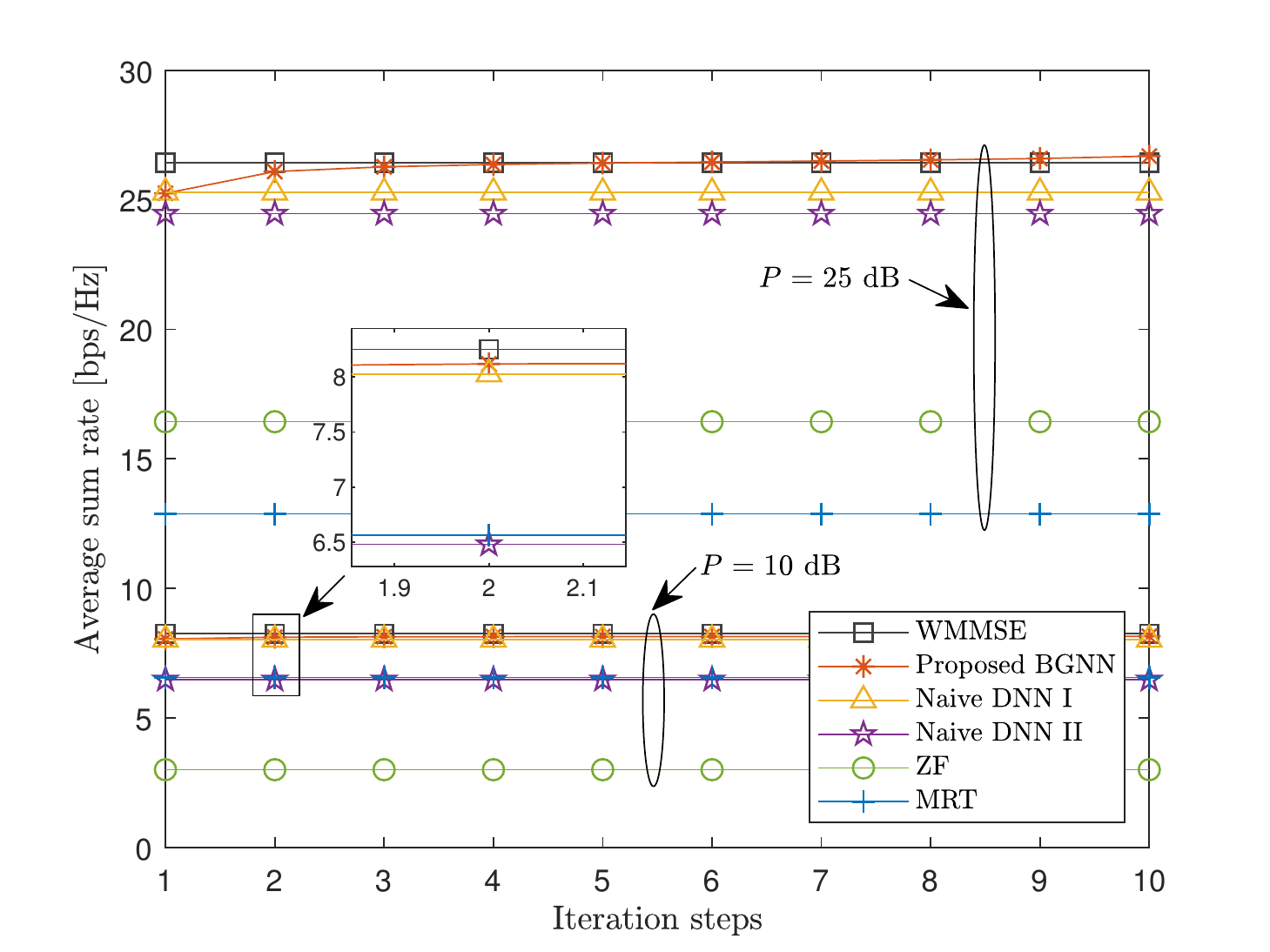}
\caption{Average sum rate versus iteration step $t$ at $N=K=8$ and $P\in\{10,25\}$ dB.}
\label{fig:fig5}
\end{figure}

Fig. \ref{fig:fig5} shows the convergence tendency of the BMP inference, i.e., the forwardpass of the BGNN. We evaluate the test sum rate performance of the BGNN trained at $T=10$ with respect to the iteration step index $t$ of Algorithm \ref{Algorithm}. The populations of antennas and users are fixed as $N=K=8$ for the test process. The BGNN outperforms the baselines only with a one-shot message-passing calculation. The performance of the BGNN gradually increases with the iteration step $t$. This verifies that the proposed BMP inference is effective to learn the convergent beamforming optimization policy. Regardless of the SNR $P$, the BGNN converges within three iteration steps. Thus, it suffices to employ three consolidations of the component DNNs \eqref{eq:eq9} in the implementation step.

\begin{table}[ht]
\centering
\caption{The average CPU running times for sum rate maximization task [sec]}\label{tab:table2}
\begin{tabular}{cccccc}
\hline \hline
\multirow{2}{*}{} & \multirow{2}{*}{Proposed BGNN} & \multirow{2}{*}{Naive DNN I} & \multirow{2}{*}{Naive DNN II} & \multicolumn{2}{c}{WMMSE} \\ \cline{5-6}
                  &                                &                              &                               & $0$ dB      & $25$ dB      \\ \hline
$N=K=4$                 & 3.7e-4                              & 2.3e-4                            & 5.5e-4                             & 4.2e-3          & 3.0e-2           \\ \hline
$N=K=6$                 & 4.1e-4                              & 4.5e-4                            & 7.7e-4                             & 8.5e-3          & 6.4e-2           \\ \hline
$N=K=8$                 & 5.1e-4                              & 7.7e-4                            & 1.1e-3                             & 1.3e-2          & 1.0e-1           \\ \hline\hline
\end{tabular}
\end{table}

Table \ref{tab:table2} compares the computational complexity of various schemes in terms of the average CPU execution time in online calculations. 
The WMMSE algorithm exhibits the worst time complexity. It generally requires more iterations as the SNR $P$ and the system size grow. In contrast, the computations of the DNN-based schemes rely on the structure of DNNs, e.g., the number of layers and neurons, and thus their CPU execution time becomes independent of $P$. Still, the time complexity of the DNN methods slightly increases with $N$ and $K$ because of the matrix inversion in the beam recovery process \eqref{eq:eq5}. The proposed BGNN method provides the lower complexity over the baseline methods thanks to the parallel computation ability. Such a complexity reduction along with the scalability is key advantages of the proposed BGNN.

\subsection{Minimum Rate Utility} \label{subsec:subsec52}

\begin{figure}
\centering
    \subfigure[$P=10$ dB, $N=4$]{
        \includegraphics[width=.33\linewidth]{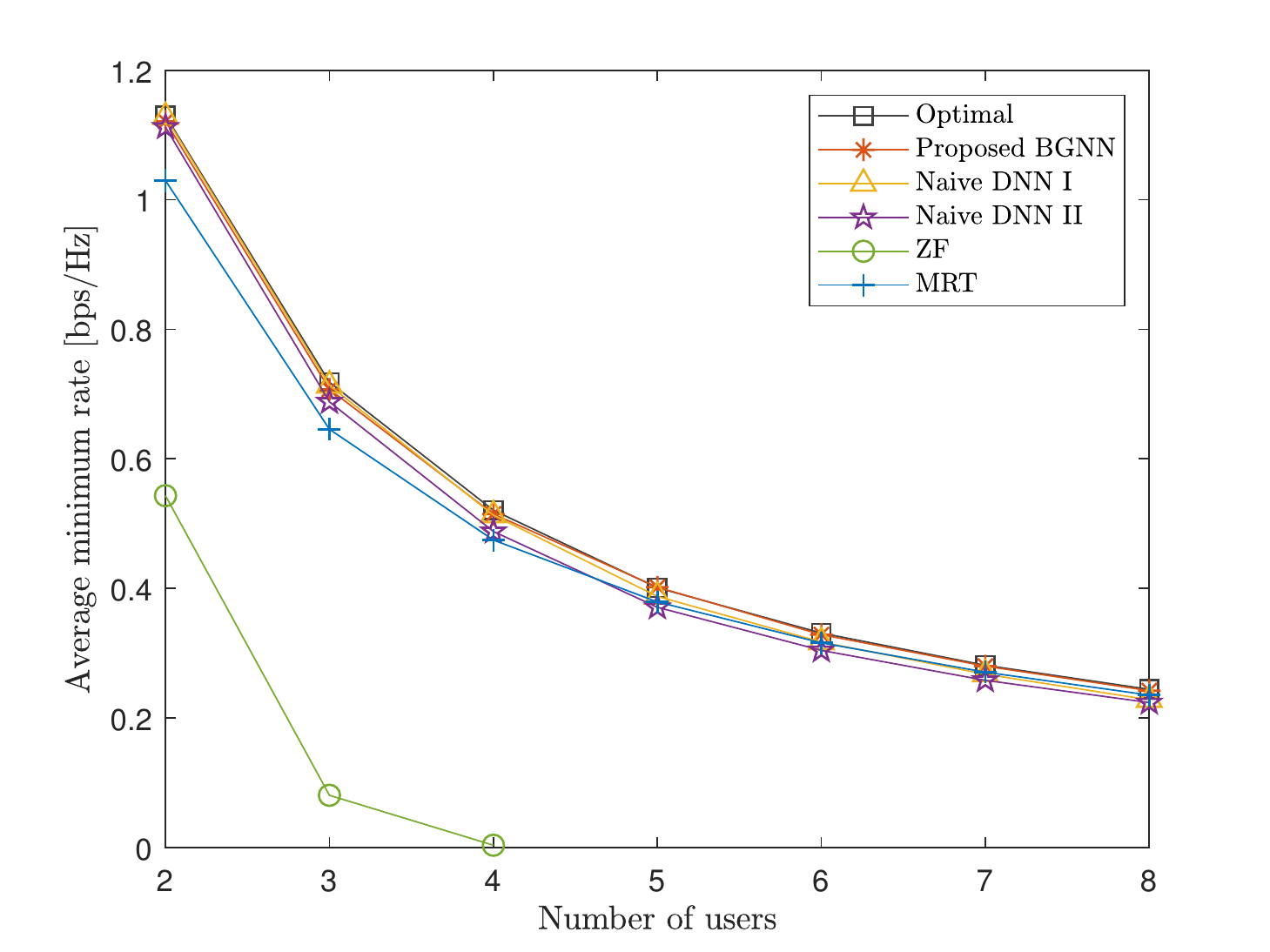}\label{fig:fig6a}
    }\hspace{-5mm}
    \subfigure[$P=10$ dB, $N=6$]{
        \includegraphics[width=.33\linewidth]{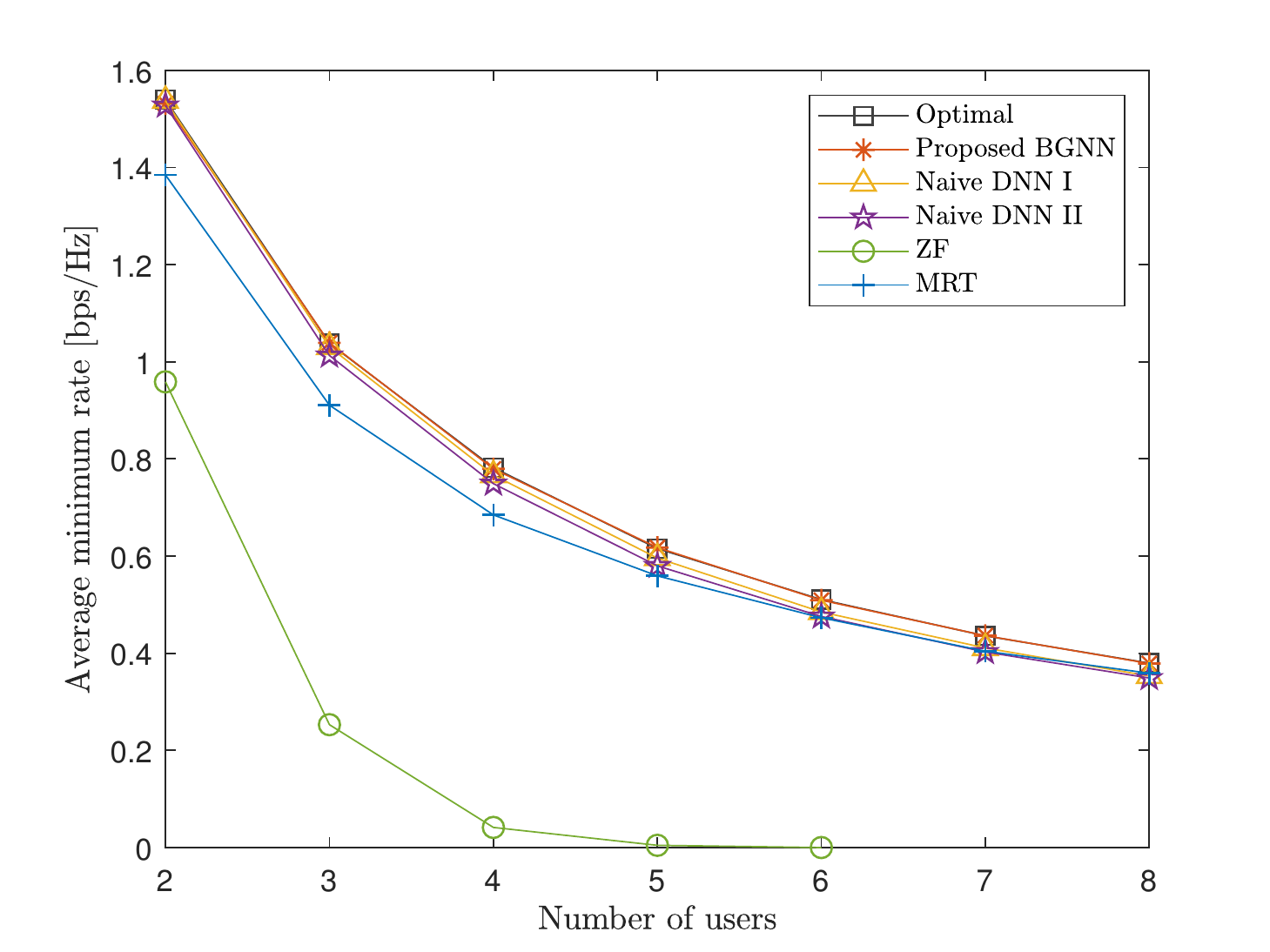}\label{fig:fig6b}
    }\hspace{-5mm}
    \subfigure[$P=10$ dB, $N=8$]{
        \includegraphics[width=.33\linewidth]{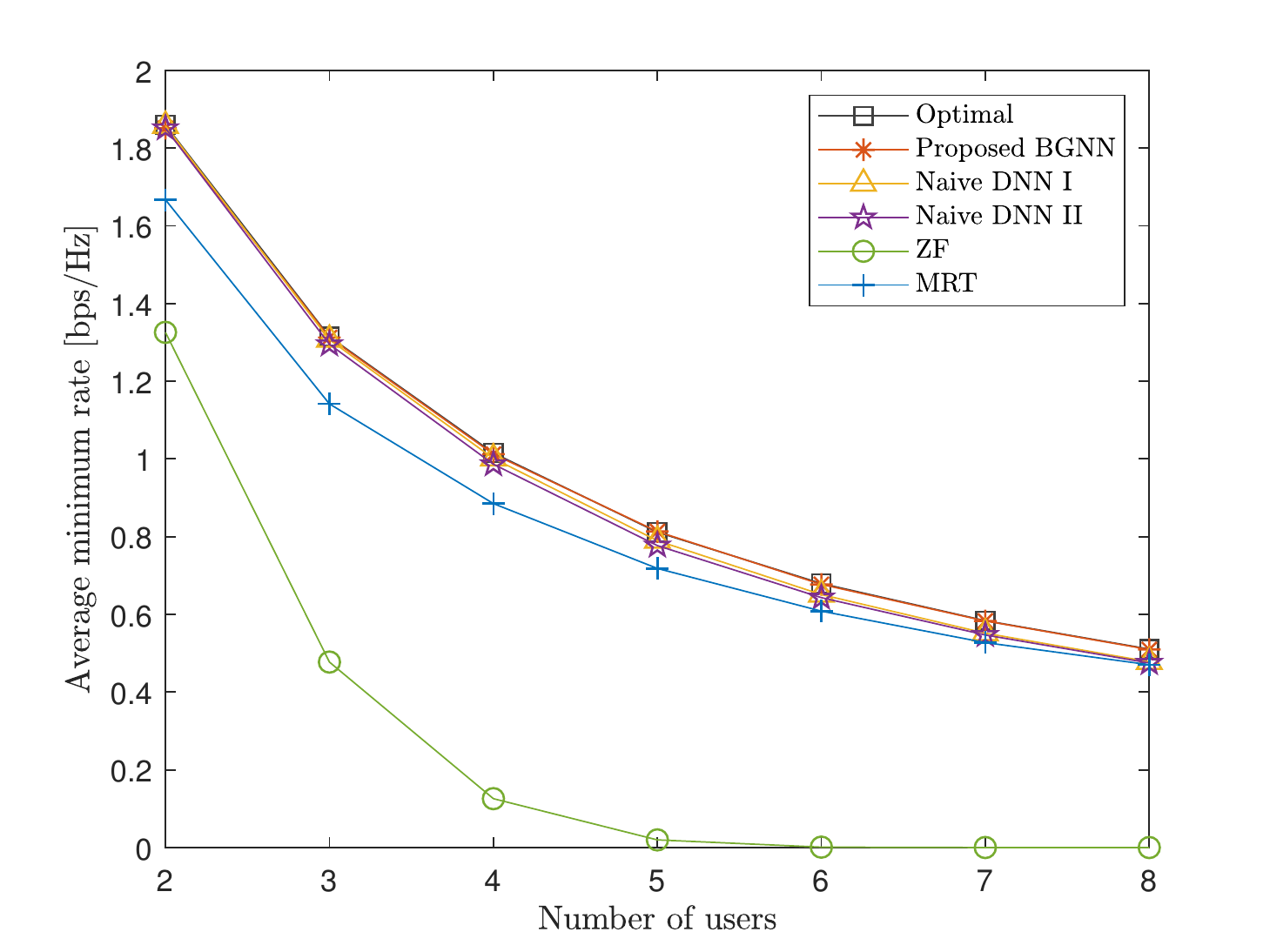}\label{fig:fig6c}
    }
    \subfigure[$P=25$ dB, $N=4$]{
        \includegraphics[width=.33\linewidth]{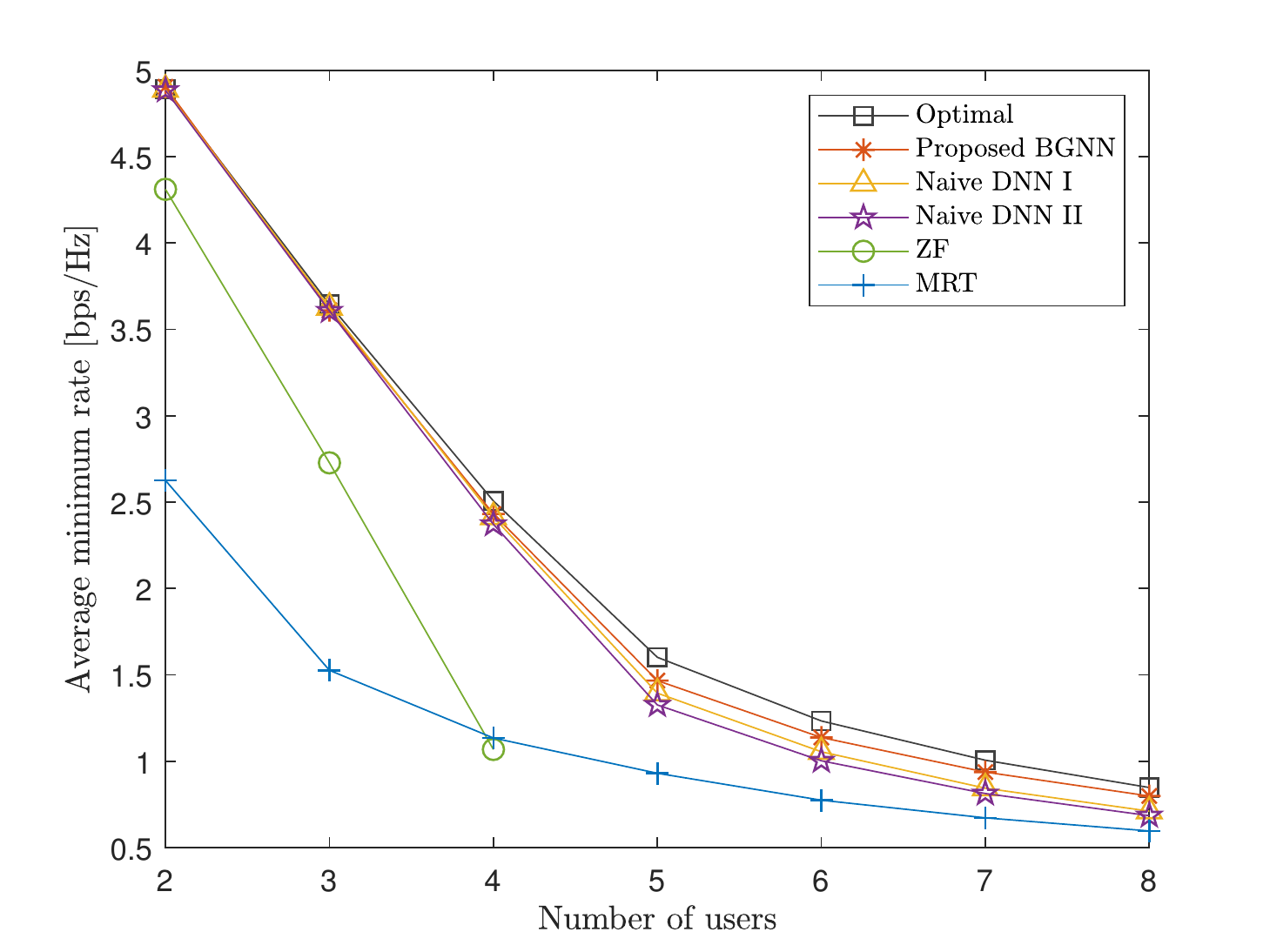}\label{fig:fig6d}
    }\hspace{-5mm}
    \subfigure[$P=25$ dB, $N=6$]{
        \includegraphics[width=.33\linewidth]{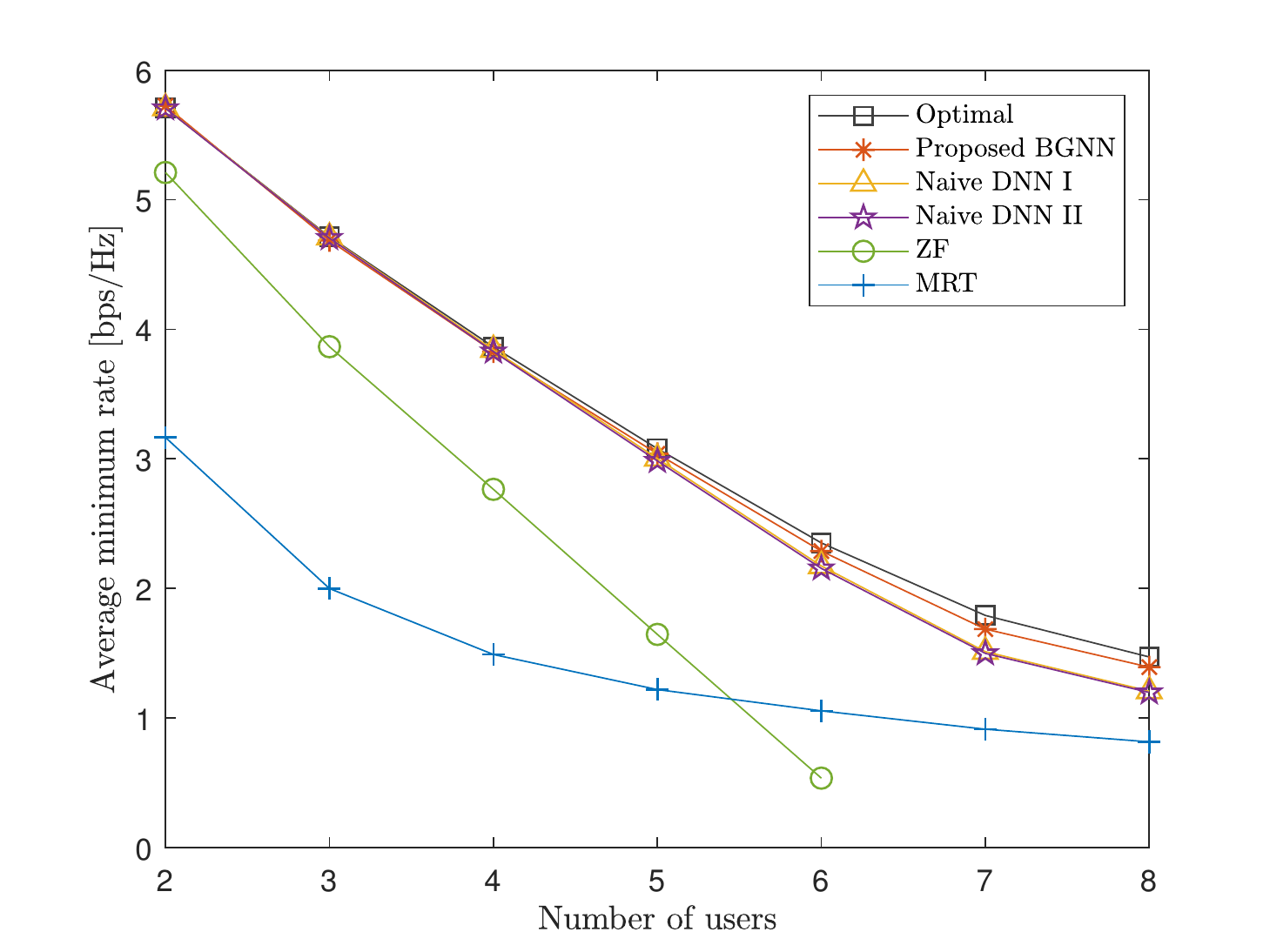}\label{fig:fig6e}
    }\hspace{-5mm}
    \subfigure[$P=25$ dB, $N=8$]{
        \includegraphics[width=.33\linewidth]{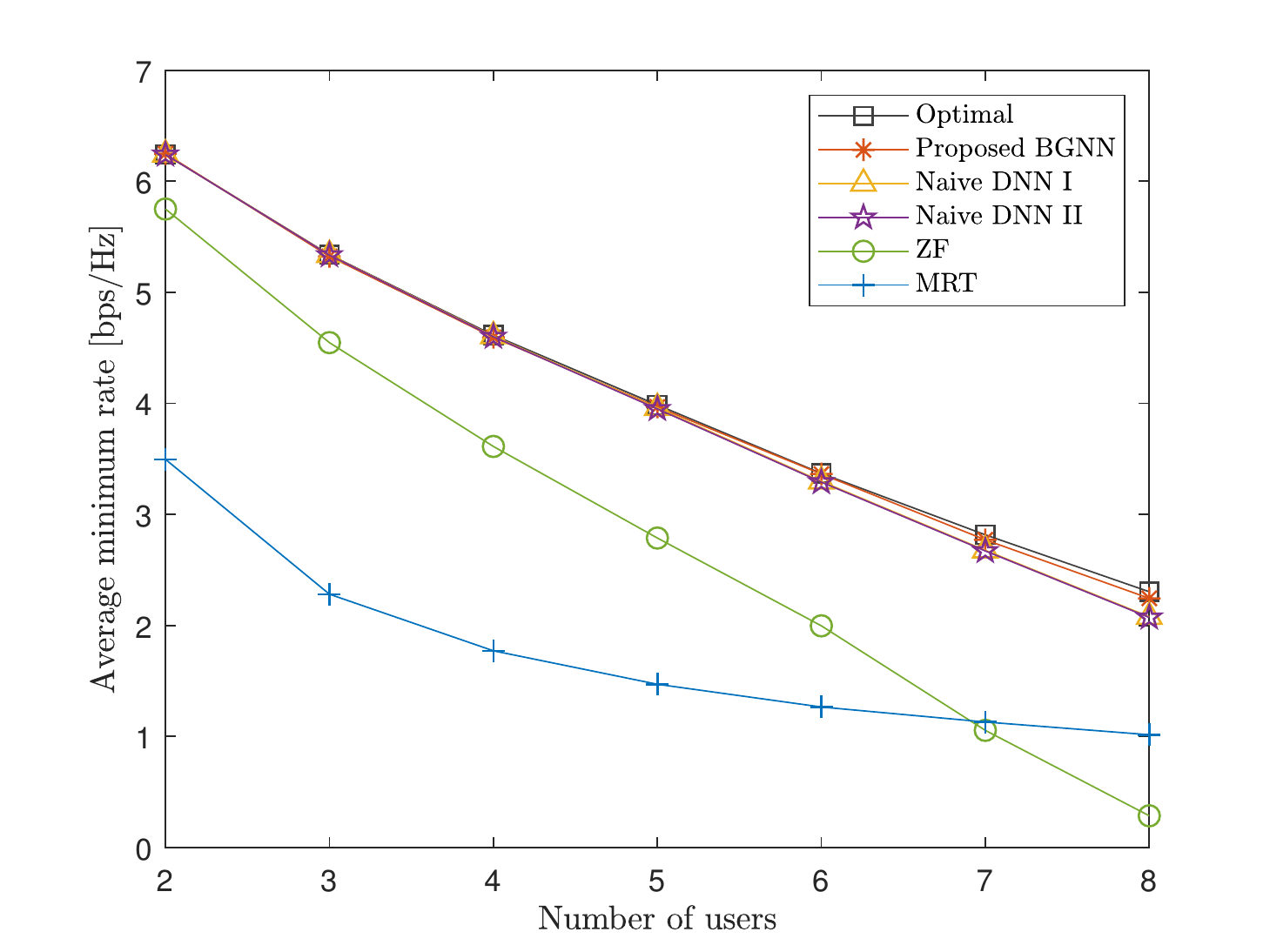}\label{fig:fig6f}
    }
    \caption{Average minimum rate performance in various network configurations.}
    \label{fig:fig6}
\end{figure}

Next, we consider the minimum rate utility function whose globally optimum solution can be found by the algorithm in \cite{Schubert:TVT04}. Fig. \ref{fig:fig6} investigates the scalability of the BGNN for the minimum rate maximization problem. We present the average minimum rate performance evaluated over the test data as a function of $K$ for $N\in\{4,6,8\}$. Similar to the sum rate utility case, the proposed scheme outperforms the naive DNN methods and achieves near-optimum performance. It is thus concluded that the BGNN can be applied to generic beamforming optimization problems.

\begin{figure}
\centering
\includegraphics[width=.45\linewidth]{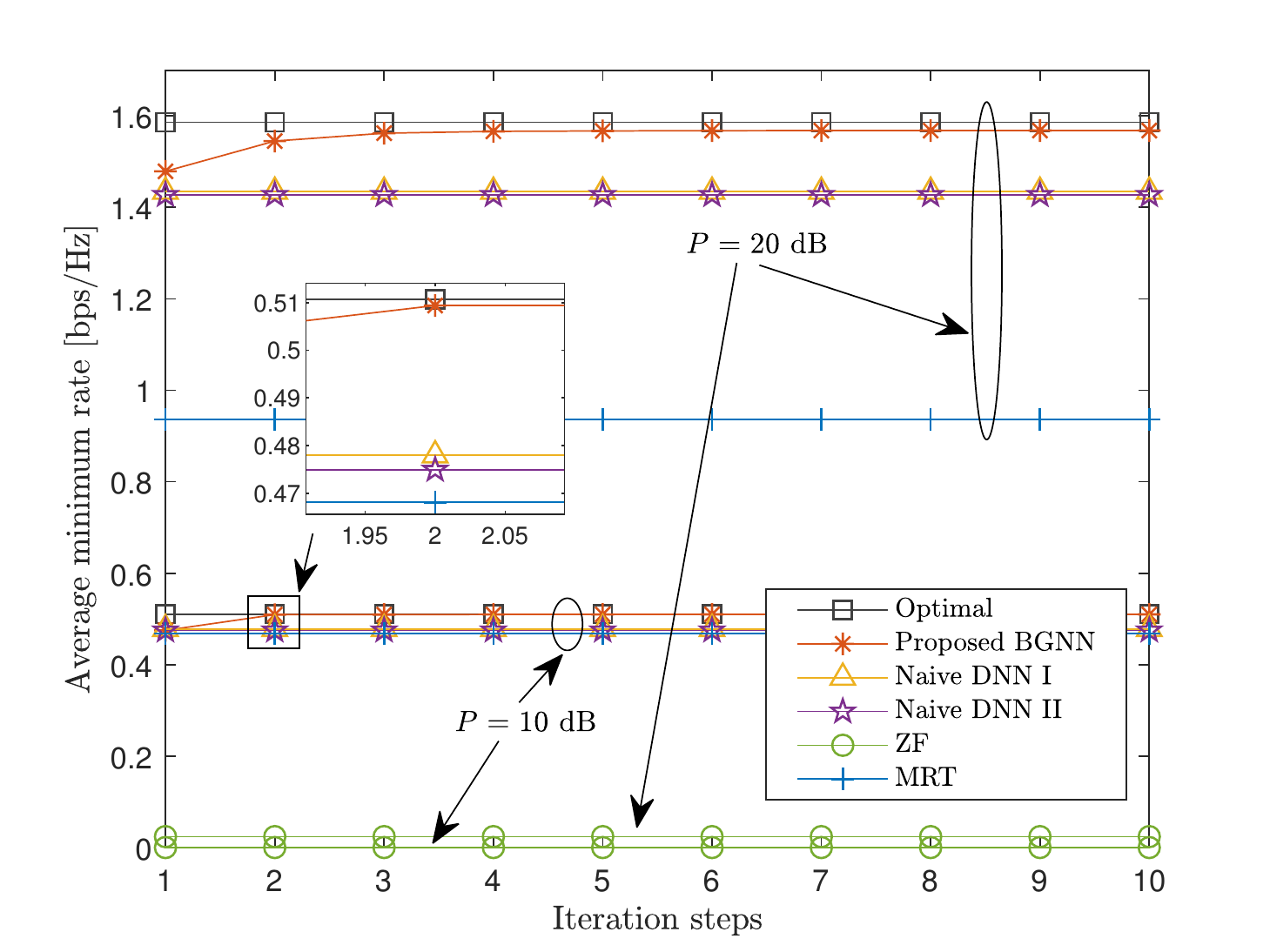}
\caption{Average minimum rate performance versus iteration step $t$ at $N=K=8$ and $P\in\{10,20\}$ dB. }
\label{fig:fig7}
\vspace{-5mm}
\end{figure}

Fig. \ref{fig:fig7} shows the convergence behavior of the average minimum rate with respect to the iteration step $t$ of the forwardpass, i.e., the BMP inference in Algorithm \ref{Algorithm}. The BGNN converges to the optimum performance within three iterations, which verifies the effectiveness of the proposed message-passing inference for the minimum rate maximization task.

\begin{figure}
\centering
\includegraphics[width=.45\linewidth]{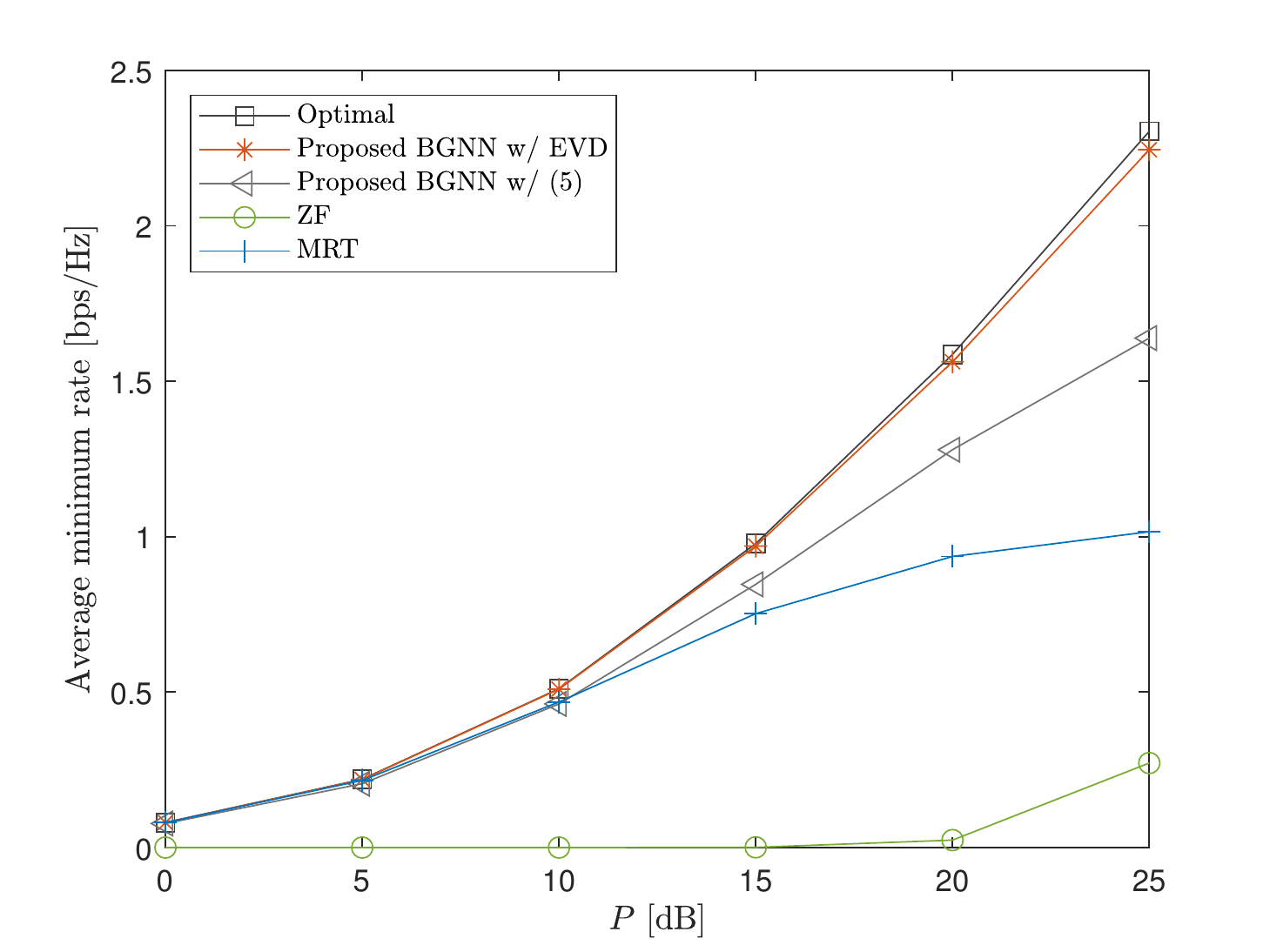}
\caption{The average minimum rate versus the SNR $P$ with $N=K=8$.}
\label{fig:fig8}
\end{figure}

In Fig. \ref{fig:fig8}, we assess the minimum rate of various schemes as a function of the SNR $P$ for $N=K=8$. To investigate the advantage of the EVD-based learning strategy provided in Section \ref{subsec:subsec42}, the BGNN is also trained with the structure in \eqref{eq:eq5} that learns both the downlink and uplink power vectors. The proposed approach shows near-optimum performance for all SNR regime. In contrast, the minimum rate performance of the BGNN trained with \eqref{eq:eq5} is degraded in the high SNR regime. This demonstrates the efficacy of the EVD-based beam recovery process and the associated training policy in Section \ref{subsec:subsec42} for the minimum rate utility.

\begin{table}[ht]
\centering
\caption{The average CPU running times for minimum rate maximization task [sec]}\label{tab:table3}
\begin{tabular}{cccccc}
\hline \hline
\multirow{2}{*}{} & \multirow{2}{*}{Proposed BGNN} & \multirow{2}{*}{Naive DNN I} & \multirow{2}{*}{Naive DNN II} & \multicolumn{2}{c}{Optimal} \\ \cline{5-6}
                  &                                &                              &                               & $0$ dB      & $25$ dB      \\ \hline
$N=K=4$                 & 8.9e-5                              & 1.0e-4                            & 1.4e-4                             & 2.0e-3          & 2.5e-3           \\ \hline
$N=K=6$                 & 1.3e-4                              & 1.9e-4                            & 2.3e-4                             & 2.9e-3          & 3.4e-3           \\ \hline
$N=K=8$                 & 2.2e-4                              & 3.2e-4                            & 3.4e-4                             & 3.6e-3          & 4.8e-3           \\ \hline\hline
\end{tabular}
\end{table}

Table \ref{tab:table3} presents the average CPU execution time of various beamforming optimization techniques. The optimal algorithm shows the worst time complexity due to intensive iterative processes. Similar to the sum rate utility case, the proposed scheme provides the best computational efficiency. A slight performance loss to the optimal algorithm is not crucial with a huge reduction of the computational complexity.

\subsection{Cell-Free MIMO Setup} \label{subsec:subsec53}

So far, we have investigated a simple multi-user network where all transmit antenna ports are co-located at the BS. In what follows, we consider a more practical cell-free MIMO setup. The BS located at the center of the cell manages separate transmit antenna ports randomly deployed on the circle of radius $30$ m. The fading coefficient $h_{k,i}$ from antenna $i$ to user $k$ is expressed as $h_{k,i}=\rho_{k,i}\tilde{h}_{k,i}$ where $\rho_{k,i}=1/(1+(d_{k,i}/d_{\text{ref}})^{\alpha})$ stands for the long-term signal attenuation with $d_{k,i}$ being the distance between antenna $i$ and user $k$, and $\tilde{h}_{k,i}$ is the small-scale fading following the zero-mean unit-variance complex Gaussian distribution.

\begin{table}[ht]
\centering
\caption{Relative minimum rate of the proposed BGNN [\%]}\label{tab:table4}
\subtable[$P=10$ dB]{
\renewcommand{\tabcolsep}{1.2mm}
\begin{tabular}{c|ccccccc}
\hline \hline
\diagbox[height=7mm]{$N$}{$K$}  & $16$ & $24$ & $32$  & $40$  & $48$  & $56$  & $64$  \\ \hline
$16$ & 99.90 & 99.93 & 99.93 & 98.80 & 98.90 & 98.90 & 99.00 \\ \hline
$24$ & 99.91 & 99.93 & 99.94 & 99.93 & 99.95 & 99.95 & 99.95 \\ \hline
$32$ & 99.94 & 99.94 & 99.94 & 99.94 & 99.95 & 99.95 & 99.95 \\ \hline
$40$ & 99.96 & 99.94 & 99.93 & 99.94 & 99.95 & 99.96 & 99.95 \\ \hline
$48$ & 99.97 & 99.95 & 99.94 & 99.94 & 99.95 & 99.96 & 99.96 \\ \hline
$56$ & 99.98 & 99.96 & 99.95 & 99.95 & 99.96 & 99.96 & 99.96 \\ \hline
$64$ & 99.99 & 99.97 & 99.95 & 99.95 & 99.96 & 99.96 & 99.96 \\ \hline \hline
\end{tabular}\label{tab:table4a}
}
\subtable[$P=25$ dB]{
\renewcommand{\tabcolsep}{1.1mm}
\begin{tabular}{c|ccccccc}
\hline \hline
\diagbox[height=7mm]{$N$}{$K$}  & $16$ & $24$ & $32$  & $40$  & $48$  & $56$  & $64$  \\ \hline
$16$ & 97.80 & 97.20 & 96.00 & 97.30 & 97.10 & 96.80 & 96.20 \\ \hline
$24$ & 99.90 & 98.50 & 98.10 & 97.30 & 97.20 & 96.90 & 96.60 \\ \hline
$32$ & 99.97 & 99.70 & 98.80 & 97.80 & 97.50 & 97.30 & 97.10 \\ \hline
$40$ & 99.99 & 99.92 & 99.60 & 98.70 & 97.90 & 97.70 & 97.70 \\ \hline
$48$ & 99.99 & 99.97 & 99.90 & 99.50 & 98.80 & 98.20 & 98.10 \\ \hline
$56$ & 99.99 & 99.99 & 99.95 & 99.90 & 99.50 & 98.90 & 98.60 \\ \hline
$64$ & 99.99 & 99.99 & 99.98 & 99.91 & 99.80 & 99.40 & 99.10 \\ \hline \hline 
\end{tabular}\label{tab:table4b}
}
\end{table}

Table \ref{tab:table4} presents the generalization ability of the BGNN in the minimum rate maximization case for $P=10$ dB (Table \ref{tab:table4a}) and $25$ dB (Table \ref{tab:table4b}). We evaluate the relative minimum rate which is defined as the achievable minimum rate of the BGNN normalized by the global optimal performance, i.e., 100 \% relative minimum rate indicates that the globally optimum performance is achieved. The BGNN trained at $N,K\in[2,8]$ is directly applied to larger cell-free MIMO configurations which were not observed in the training step. Similar to the sum-rate maximization case shown in Fig. 3, the proposed BGNN works well in unseen systems as large as $N=K=64$. This result confirms that the proposed approach can effectively handle arbitrary distributed MIMO networks having massive populations of antennas and users.

\begin{figure}
\centering
    \subfigure[$P=10$ dB]{
        \includegraphics[width=.35\linewidth]{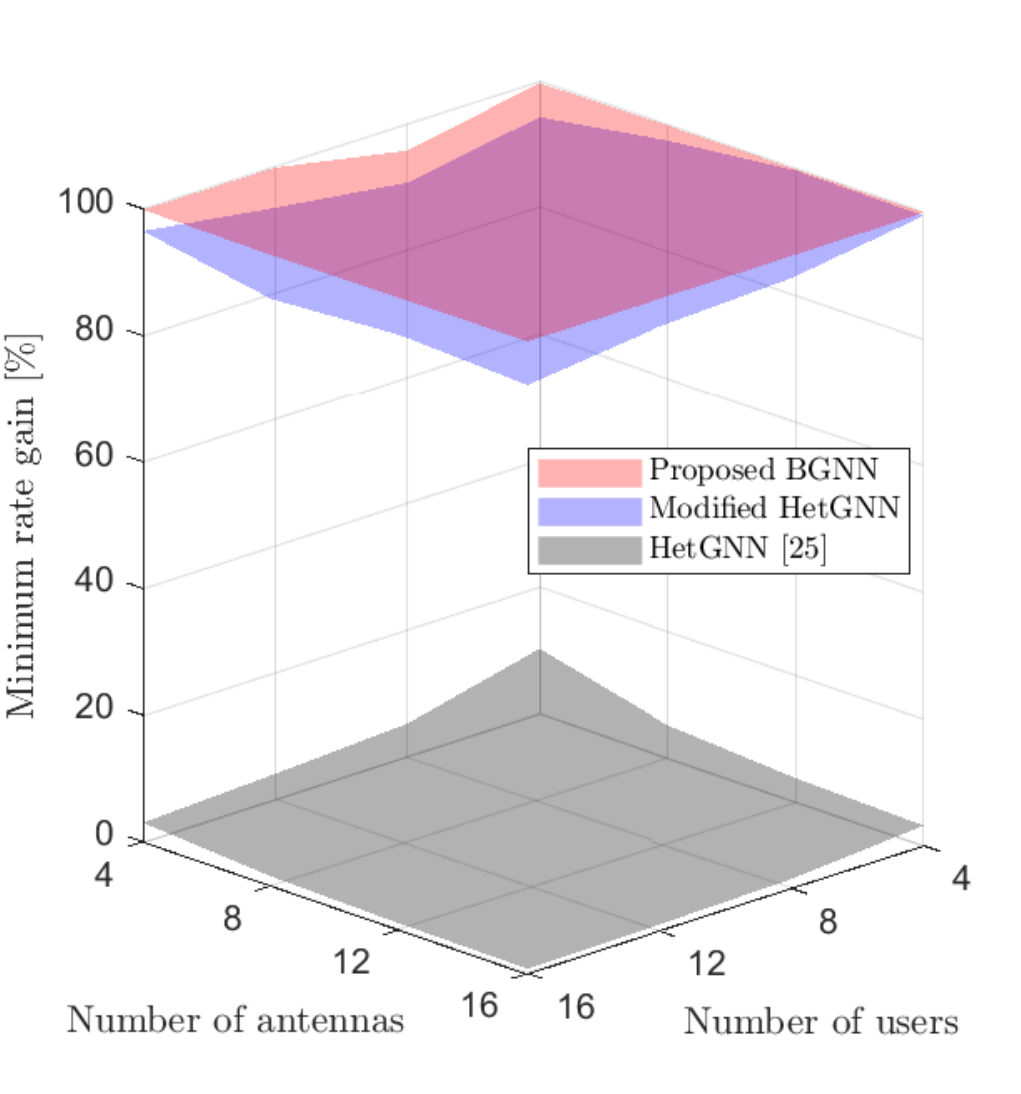}\label{fig:fig9a}
    }\hspace{5mm}
    \subfigure[$P=25$ dB]{
        \includegraphics[width=.35\linewidth]{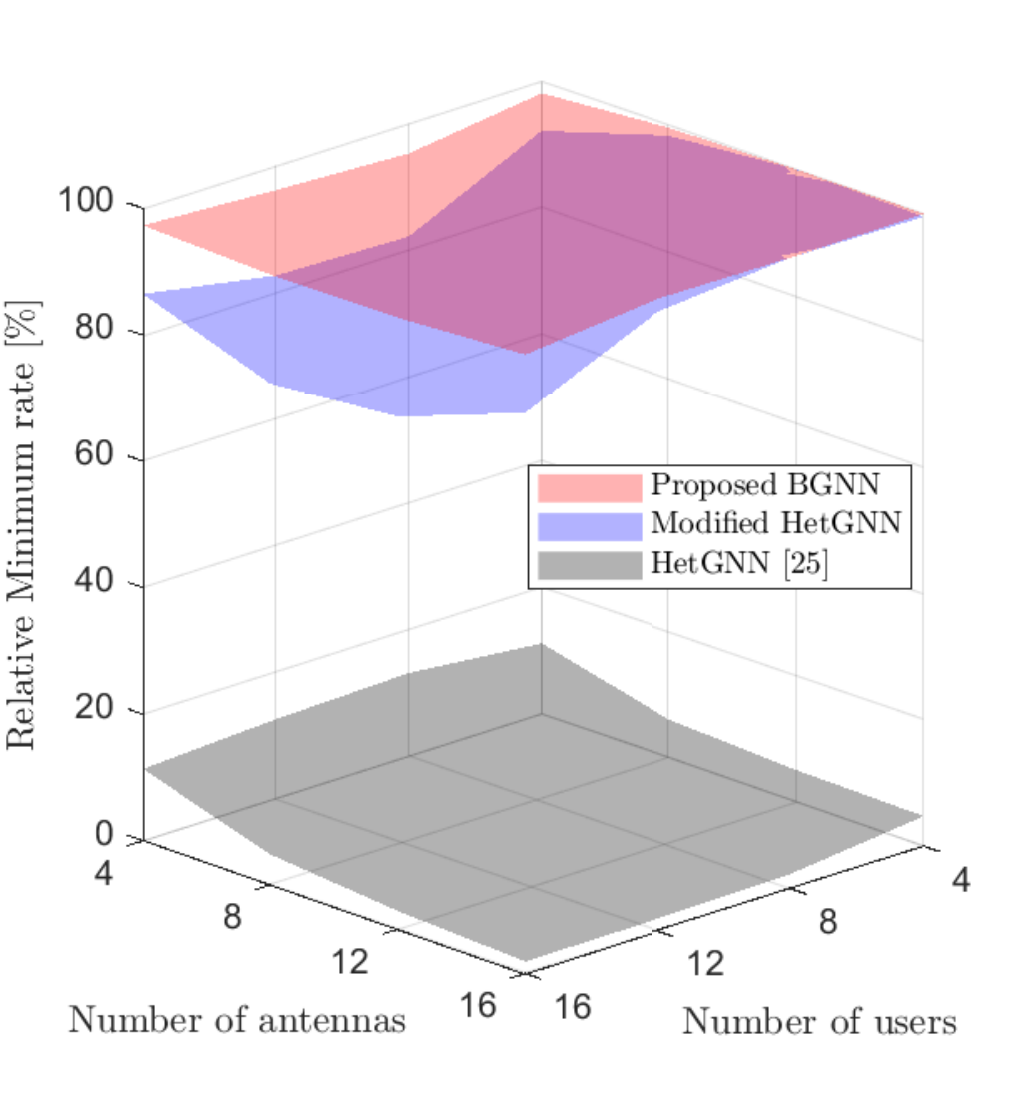}\label{fig:fig9b}
    }
    \caption{Relative minimum rate performance with respect to $N$ and $K$.}
    \label{fig:fig9}
\end{figure}

Fig. \ref{fig:fig9} compares the relative minimum rate performance achieved by the proposed BGNN and conventional HetGNN method \cite{Guo:TWC22} with respect to the numbers of antennas and users for $P\in\{10,25\}$ dB. As in \eqref{eq:eqHetGNN1c}, the HetGNN \cite{Guo:TWC22} straightforwardly outputs the beamforming vectors. As discussed in Section \ref{subsec:subsec22}, 
such a \textit{direct learning} policy has been proven to be inefficient to handle complex-valued optimization problems \cite{Xia:TC20, Kim:WCL20}. Therefore, likewise the proposed BGNN, we also examine the modified HetGNN that yields the beam feature outputs $\{q_{k}:\forall k\}$ at the antenna vertices. From the figure, we can see that the conventional HetGNN fails to achieve a good performance in all simulated network configurations. The modified HetGNN significantly improves the minimum rate performance, implying that the proposed feature learning strategy is essential for the beamforming optimization problems. Nevertheless, the modified HetGNN still performs worse than the proposed BGNN that is specialized to the beamforming optimization problem. This validates the effectiveness of the proposed BMP inference for retrieving the sufficient statistics of the optimum beamforming vectors.

\section{Conclusion} \label{sec:sec6}
This paper has developed a BGNN framework for addressing scalable beamforming tasks having randomly scheduled antennas and users. We have interpreted the MU-MISO system as a weighted bipartite graph where two disjoint vertex sets, respectively consisting of antennas and users, interact with each other through edges weighted by channel fading coefficients. The BMP inference has been derived which generates convergent beamforming solutions over arbitrary bipartite graphs. A tractable realization of the BMP inference, which is referred to as the BGNN, has been proposed. It comprises three types of reusable component DNNs characterizing calculation processes of antenna vertices and user vertices. Changes in antenna and user configurations are easily reflected in the BGNN by removing or adding the corresponding DNN modules. This brings the dimensionality-invariant computation nature so that the BGNN can be universally applied to arbitrary communication systems. Such a generalization property is further improved by using the proposed training policy that includes a number of random bipartite graphs into training samples. The effectiveness of the proposed framework has been examined in the sum rate and minimum rate maximization problems. Numerical results have verified the scalability of the proposed approach in larger systems that are unseen in the training~step.

\begin{appendices}
\section{Derivation of BMP Inference}\label{appendix}
Without the loss of the optimality, the beam feature $\mathbf{s}_{k}$ of user $k\in\mathcal{K}$ can be optimally computed by using the channel matrix $\mathbf{H}=\{h_{k,i}:\forall k,i\}$ and the other users' beam features $\mathbf{s}_{l}$, $\forall l\neq k$. The objective function of \eqref{eq:eq4} exhibits the permutation-invariant property in which the ordering of user indices has no impacts on its calculation. Based on this intuition, we can conclude that the computation process of the optimal beam feature has the identical structure at all users. Such a shared beam feature optimization process is denoted by an operator $\mathcal{D}(\cdot)$. Then, $\mathbf{s}_{k}$ can be expressed as
\begin{align} \label{eq:eq16}
\mathbf{s}_{k} = \mathcal{D}\big(\{\mathbf{s}_{l}:\forall l\neq k\},\{h_{l,j}:\forall l, j\}\big).
\end{align}
The requirement of other beam features $\{\mathbf{s}_{l}:\forall l\neq k\}$ in \eqref{eq:eq16} triggers fixed point iterations among $\mathbf{s}_{k}$, $\forall k\in\mathcal{K}$. For this reason, users need to update their beam features repeatedly by exchanging previous decisions. We rewrite the decision process in \eqref{eq:eq16} as
\begin{subequations}\label{eq:eq17}
\begin{align}
\mathbf{s}_{k}^{[t]} & =  \mathcal{D}\big(\mathbf{s}_{k}^{[t-1]}, \{\mathbf{s}_{l}^{[t-1]}:\forall l\neq k\},\{h_{l,j}:\forall l, j\}\big) \label{eq:eq17a} \\
& =  \mathcal{D}\big(\{\mathbf{b}_{j,k}^{[t]}:\forall j\}, \{\mathbf{b}_{j,l}^{[t]}:\forall l \neq k, \forall j\}\big), \label{eq:eq17b}
\end{align}
\end{subequations}
where \eqref{eq:eq17a} introduces the past knowledge about other users $\{\mathbf{s}_{l}^{[t-1]}:\forall l\neq k\}$ as well as the internal decision $\mathbf{s}_{k}^{[t-1]}$ as input data and $\mathbf{b}_{i,k}^{[t]}$ in \eqref{eq:eq17b} is defined as a tuple $\mathbf{b}_{i,k}^{[t]} = (\mathbf{s}_{k}^{[t-1]},h_{k,i})$.

The recursion equation in \eqref{eq:eq17} provides an operator $\mathcal{M}(\cdot)$ for generating $\mathbf{b}_{i,k}^{[t]}$~as
\begin{align} \label{eq:eq18}
\mathbf{b}_{i,k}^{[t]} & = \big(\mathbf{s}_{k}^{[t-1]},h_{k,i}\big) = \mathcal{M}\big(\{\mathbf{b}_{j,k}^{[t-1]}:\forall j\}, \{\mathbf{b}_{j,l}^{[t-1]}:\forall\ l\neq k, \forall j\}, h_{k,i}\big). 
\end{align}
Denoting $\mathbf{c}_{k,i}^{[t]} = \{\mathbf{b}_{j,k}^{[t-1]}:\forall j\neq i\}$ simplifies \eqref{eq:eq18} into
\begin{subequations}\label{eq:eq19}
\begin{align}
\mathbf{b}_{i,k}^{[t]}
& = \mathcal{M}\big(\mathbf{b}_{i,k}^{[t-1]}, \{\mathbf{c}_{l,i}^{[t]}:\forall\ l\}, \{\mathbf{b}_{i,l}^{[t-1]}:\forall\ l\neq k\}, h_{k,i}\big)\\
& = \mathcal{M}\big(\mathbf{m}_{i,k}^{[t-1]}, \{\mathbf{c}_{l,i}^{[t]}:\forall\ l\}, h_{k,i}\big),
\end{align}
\end{subequations}
where $\mathbf{m}_{i,k}^{[t]}\triangleq(\mathbf{b}_{i,k}^{[t]}, \{\mathbf{b}_{i,l}^{[t]}:\forall l\neq k\})$. With a proper pooling operator $\mathcal{P}(\cdot)$ in \eqref{eq:eq7c}, \eqref{eq:eq19} is recast to the antenna message generation in \eqref{eq:eq6b}. Plugging $\mathbf{m}_{i,k}^{[t]}=(\mathbf{b}_{i,k}^{[t]}, \{\mathbf{b}_{i,l}^{[t]}:\forall l\neq k\})$ and \eqref{eq:eq7d} into \eqref{eq:eq17b} leads to the simplified decision operation in \eqref{eq:eq6c}.

In addition, we employ an operator $\mathcal{C}(\cdot)$ to compute the information vector $\mathbf{c}_{k,i}^{[t]}=\{\mathbf{b}_{j,k}^{[t-1]}:\forall j\neq i\}$ as
\begin{subequations}\label{eq:eq20}
\begin{align}
\mathbf{c}_{k,i}^{[t]} &= \mathcal{C}\big(\mathbf{b}_{i,k}^{[t-1]}, \{\mathbf{b}_{j,k}^{[t-1]}:\forall j \}\big)\\
& = \mathcal{C}\big(\mathbf{s}_{k}^{[t-1]}, \mathbf{b}_{k}^{[t-1]}, h_{k,i}\big),\label{eq:eq20b}
\end{align}
\end{subequations}
where \eqref{eq:eq20b} comes from the definition $\mathbf{b}_{i,k}^{[t]} = (\mathbf{s}_{k}^{[t-1]},h_{k,i})$ and the pooling operation in \eqref{eq:eq7a}. All processes of the BMP inference in \eqref{eq:eq6} have been obtained. \qqed

\end{appendices}

\renewcommand{\baselinestretch}{1.23}

\bibliographystyle{IEEEtran}

\bibliography{arxiv}

\begin{thebibliography}{10}
\providecommand{\url}[1]{#1}
\csname url@samestyle\endcsname
\providecommand{\newblock}{\relax}
\providecommand{\bibinfo}[2]{#2}
\providecommand{\BIBentrySTDinterwordspacing}{\spaceskip=0pt\relax}
\providecommand{\BIBentryALTinterwordstretchfactor}{4}
\providecommand{\BIBentryALTinterwordspacing}{\spaceskip=\fontdimen2\font plus
\BIBentryALTinterwordstretchfactor\fontdimen3\font minus
  \fontdimen4\font\relax}
\providecommand{\BIBforeignlanguage}[2]{{%
\expandafter\ifx\csname l@#1\endcsname\relax
\typeout{** WARNING: IEEEtran.bst: No hyphenation pattern has been}%
\typeout{** loaded for the language `#1'. Using the pattern for}%
\typeout{** the default language instead.}%
\else
\language=\csname l@#1\endcsname
\fi
#2}}
\providecommand{\BIBdecl}{\relax}
\BIBdecl

\bibitem{Bjornson:SPM14}
E.~{Björnson}, M.~{Bengtsson}, and B.~{Ottersten}, ``Optimal multiuser
  transmit beamforming: A difficult problem with a simple solution structure
  [lecture notes],'' \emph{IEEE Signal Process. Mag.}, vol.~31, no.~4, pp.
  142--148, Jul. 2014.

\bibitem{Shi:TSP11}
Q.~{Shi}, M.~{Razaviyayn}, Z.~{Luo}, and C.~{He}, ``An iteratively weighted
  {MMSE} approach to distributed sum-utility maximization for a {MIMO}
  interfering broadcast channel,'' \emph{IEEE Trans. Signal Process.}, vol.~59,
  no.~9, pp. 4331--4340, Sept. 2011.

\bibitem{Christensen:TWC08}
S.~S. {Christensen}, R.~{Agarwal}, E.~{De Carvalho}, and J.~M. {Cioffi},
  ``Weighted sum-rate maximization using weighted {MMSE} for {MIMO-BC}
  beamforming design,'' \emph{IEEE Trans. Wireless Commun.}, vol.~7, no.~12,
  pp. 4792--4799, Dec. 2008.

\bibitem{Schubert:TVT04}
M.~{Schubert} and H.~{Boche}, ``Solution of the multiuser downlink beamforming
  problem with individual {SINR} constraints,'' \emph{IEEE Trans. Veh.
  Technol.}, vol.~53, no.~1, pp. 18--28, Jan. 2004.

\bibitem{Boche:VTC02}
H.~{Boche} and M.~{Schubert}, ``A general duality theory for uplink and
  downlink beamforming,'' in \emph{Proc. IEEE Veh. Technol. Conf. (VTC)}, 2002,
  pp. 87--91.

\bibitem{Zappone:TC19}
A.~{Zappone}, M.~{Di Renzo}, and M.~{Debbah}, ``Wireless networks design in the
  era of deep learning: Model-based, {AI}-based, or both?'' \emph{IEEE Trans.
  Commun.}, vol.~67, no.~10, pp. 7331--7376, Oct. 2019.

\bibitem{Zhang:CST19}
C.~{Zhang}, P.~{Patras}, and H.~{Haddadi}, ``Deep learning in mobile and
  wireless networking: A survey,'' \emph{IEEE Commun. Surv. Tutor.}, vol.~21,
  no.~3, pp. 2224--2287, 3rf Quart. 2019.

\bibitem{Sun:TSP18}
H.~{Sun}, X.~{Chen}, Q.~{Shi}, M.~{Hong}, X.~{Fu}, and N.~D. {Sidiropoulos},
  ``Learning to optimize: Training deep neural networks for interference
  management,'' \emph{IEEE Trans. Signal Process.}, vol.~66, no.~20, pp.
  5438--5453, Oct. 2018.

\bibitem{Xia:TC20}
W.~{Xia}, G.~{Zheng}, Y.~{Zhu}, J.~{Zhang}, J.~{Wang}, and A.~P. {Petropulu},
  ``A deep learning framework for optimization of {MISO} downlink
  beamforming,'' \emph{IEEE Trans. Commun.}, vol.~68, no.~3, pp. 1866--1880,
  Mar. 2020.

\bibitem{Kim:WCL20}
J.~{Kim}, H.~{Lee}, S.~E. {Hong}, and S.~H. {Park}, ``Deep learning methods for
  universal {MISO} beamforming,'' \emph{IEEE Wireless Commun. Lett.}, vol.~9,
  no.~11, pp. 1894--1898, Nov. 2020.

\bibitem{Kim:CL21}
J.~{Kim}, H.~{Lee}, and S.~H. {Park}, ``Learning robust beamforming for {MISO}
  downlink systems,'' \emph{IEEE Commun. Lett.}, vol.~25, no.~6, pp.
  1916--1920, Jun. 2021.

\bibitem{Huang:TVT20}
H.~{Huang}, Y.~{Peng}, J.~{Yang}, W.~{Xia}, and G.~{Gui}, ``Fast beamforming
  design via deep learning,'' \emph{IEEE Trans. Veh. Technol.}, vol.~69, no.~1,
  pp. 1065--1069, Jan. 2020.

\bibitem{Liang:TC20}
F.~{Liang}, C.~{Shen}, W.~{Yu}, and F.~{Wu}, ``Towards optimal power control
  via ensembling deep neural networks,'' \emph{IEEE Trans. Commun.}, vol.~68,
  no.~3, pp. 1760--1776, Mar. 2020.

\bibitem{Kim:ACSSC18}
M.~{Kim}, P.~d.~{Kerret}, and D.~{Gesbert}, ``Learning to cooperate in
  decentralized wireless networks,'' in \emph{Proc. Asilomar Conf. Signals,
  Syst. Comput. (ACSSC)}, Mar. 2018, pp. 281--285.

\bibitem{Lee:JSAC19}
H.~{Lee}, S.~H. {Lee}, and T.~Q.~S. {Quek}, ``Deep learning for distributed
  optimization: Applications to wireless resource management,'' \emph{IEEE J.
  Sel. Areas Commun.}, vol.~37, no.~10, pp. 2251--2266, Oct. 2019.

\bibitem{Lee:TWC21.1}
H.~{Lee}, J.~{Kim}, and S.~H. {Park}, ``Learning optimal fronthauling and
  decentralized edge computation in fog radio access networks,'' \emph{IEEE
  Trans. Wireless Commun.}, vol.~20, no.~9, pp. 5599--5612, Sept. 2021.

\bibitem{Wu:TNNLS21}
Z.~{Wu}, S.~{Pan}, F.~{Chen}, G.~{Long}, C.~{Zhang}, and P.~S. {Yu}, ``A
  comprehensive survey on graph neural networks,'' \emph{IEEE Trans. Neural
  Netw. Learn. Syst.}, vol.~32, no.~1, pp. 4--24, Jan. 2021.

\bibitem{Zhou:arXiv18}
J.~{Zhou}, G.~{Cui}, Z.~{Zhang}, C.~{Yang}, Z.~{Liu}, and M.~{Sun}, ``Graph
  neural networks: {A} review of methods and applications,'' \emph{arXiv
  preprint arXiv:1812.08434}, 2018, [Online] Available:
  https://arxiv.org/abs/1812.08434.

\bibitem{Shen:JSAC21}
Y.~{Shen}, Y.~{Shi}, J.~{Zhang}, and K.~B. {Letaief}, ``Graph neural networks
  for scalable radio resource management: Architecture design and theoretical
  analysis,'' \emph{IEEE J. Sel. Areas Commun.}, vol.~39, no.~1, pp. 101--115,
  Jan. 2021.

\bibitem{Eisen:TSP20}
M.~{Eisen} and A.~{Ribeiro}, ``Optimal wireless resource allocation with random
  edge graph neural networks,'' \emph{IEEE Trans. Signal Process.}, vol.~68,
  pp. 2977--2991, Jun. 2020.

\bibitem{Lee:TWC21}
H.~{Lee}, S.~H. {Lee}, and T.~Q.~S. {Quek}, ``Learning autonomy in management
  of wireless random networks,'' \emph{IEEE Trans. Wireless Commun.}, vol.~20,
  no.~12, pp. 8039--8053, Dec. 2021.

\bibitem{Chowdhury:arXiv20}
A.~{Chowdhury}, G.~{Verma}, C.~{Rao}, A.~{Swami}, and S.~{Segarra}, ``Unfolding
  {WMMSE} using graph neural networks for efficient power allocation,''
  \emph{IEEE Trans. Wireless Commun.}, vol.~20, no.~9, pp. 6004--6017, Sept.
  2021.

\bibitem{Yang:CL20}
Y.~{Yang}, S.~{Zhang}, F.~{Gao}, J.~{Ma}, and O.~A. {Dobre}, ``Graph neural
  network-based channel tracking for massive {MIMO} networks,'' \emph{IEEE
  Commun. Lett.}, vol.~24, no.~8, pp. 1747--1751, Aug. 2020.

\bibitem{Jiang:JSAC21}
T.~{Jiang}, H.~V. {Cheng}, and W.~{Yu}, ``Learning to reflect and to beamform
  for intelligent reflecting surface with implicit channel estimation,''
  \emph{IEEE J. Sel. Areas Commun.}, vol.~39, no.~7, pp. 1931--1945, May 2021.

\bibitem{Guo:TWC22}
J.~{Gua} and C.~{Yang}, ``Learning power allocation for multi-cell-multi-user
  systems with heterogeneous graph neural networks,'' \emph{IEEE Trans.
  Wireless Commun.}, vol.~21, no.~2, pp. 884--897, Feb. 2022.

\bibitem{Zhang:arXiv21}
X.~{Zhang}, H.~{Zhao}, J.~{Xiong}, L.~{Zhou}, and J.~{Wei}, ``Scalable power
  control/beamforming in heterogeneous wireless networks with graph neural
  networks,'' \emph{arXiv preprint arXiv:2104.05463}, 2021, [Online] Available:
  https://arxiv.org/abs/2104.05463.

\bibitem{Zhang:ACM19}
C.~{Zhang}, D.~{Song}, C.~{Huang}, A.~{Swami}, and N.~V. {Chawla},
  ``Heterogeneous graph neural network,'' in \emph{Proc. 25th ACM SIGKDD Int.
  Conf. Knowl. Discov. Data Mining}, 2019, pp. 793--803.

\bibitem{Ngo:TWC17}
H.~Q. {Ngo}, A.~{Ashikhmin}, H.~{Yang}, E.~G. {Larsson}, and T.~L. {Marzetta},
  ``Cell-free massive {MIMO} versus small cells,'' \emph{IEEE Trans. Wireless
  Commun.}, vol.~16, no.~3, pp. 1834--1850, Mar. 2017.

\bibitem{Nayebi:TWC17}
E.~{Nayebi}, A.~{Ashikhmin}, T.~L. {Marzetta}, H.~{Yang}, and B.~D. {Rao},
  ``Precoding and power optimization in cell-free massive {MIMO} systems,''
  \emph{IEEE Trans. Wireless Commun.}, vol.~16, no.~7, pp. 4445--4459, Jul.
  2017.

\bibitem{Khalili:TWC20}
A.~{Khalili}, M.~R. {Mili}, M.~{Rasti}, S.~{Parsaeefard}, and D.~W.~K. {Ng},
  ``Antenna selection strategy for energy efficiency maximization in uplink
  {OFDMA} networks: {A} multi-objective approach,'' \emph{IEEE Trans. Wireless
  Commun.}, vol.~19, no.~1, pp. 595--609, Jan. 2020.

\bibitem{Bai:TIT09}
D.~{Bai}, P.~{Mitran}, S.~S. {Ghassemzadeh}, R.~R. {Miller}, and V.~{Tarokh},
  ``Rate of channel hardening of antenna selection diversity schemes and its
  implication on scheduling,'' \emph{IEEE Trans. Inf. Theory}, vol.~55, no.~10,
  pp. 4353--4365, Oct. 2009.

\bibitem{MZaheer:17}
M.~{Zaheer}, S.~{Kottur}, S.~{Ravanbakhsh}, B.~{Poczos}, R.~R. {Salakhutdinov},
  and A.~J. {Smola}, ``Deep sets,'' in \emph{Proc. Adv. Neural Inf. Process.
  Syst. (NeurIPS)}, Dec. 2017, pp. 3391--3401, [Online] Available:
  https://arxiv.org/abs/1703.06114.

\bibitem{Hornik:NN89}
K.~Hornik, M.~Stinchcombe, and H.~White, ``Multilayer feedforward networks are
  universal approximators,'' \emph{Neural Netw.}, vol.~2, no.~5, pp. 359--366,
  Jan. 1989.

\bibitem{He:CVPR16}
K.~{He}, X.~{Zhang}, S.~{Ren}, and J.~{Sun}, ``Deep residual learning for image
  recognition,'' in \emph{Proc. IEEE Conf. Comput. Vis. Pattern Recognit.
  (CVPR)}, Jun. 2016, pp. 770--778.

\bibitem{Huang:CVPR17}
G.~{Huang}, Z.~{Liu}, L.~V.~D. {Maaten}, and K.~Q. {Weinberger}, ``Densely
  connected convolutional networks,'' in \emph{Proc. IEEE Conf. Comput. Vis.
  Pattern Recognit. (CVPR)}, Jul. 2017, pp. 4700--4708.

\bibitem{Kingma:ICLR15}
D.~Kingma and J.~Ba, ``Adam: A method for stochastic optimization,'' in
  \emph{Proc. Int. Conf. Learn. Represent. (ICLR)}, 2015, pp. 1--15.

\bibitem{Srivastava:14}
N.~{Srivastava}, G.~E. {Hinton}, A.~{Krizhevsky}, I.~{Sutskever}, and
  R.~{Salakhutdinov}, ``Dropout: {A} simple way to prevent neural networks from
  overfitting,'' \emph{J. Mach. Learn. Res.}, vol.~15, no.~1, pp. 1929--1958,
  2014.

\bibitem{Ruiz:TSP20}
L.~{Ruiz}, F.~{Gama}, and A.~{Ribeiro}, ``Gated graph recurrent neural
  networks,'' \emph{IEEE Trans. Signal Process.}, vol.~68, pp. 6303--6318, Oct.
  2020.

\bibitem{Segarra:TSP17}
S.~{Segarra}, A.~G. {Marques}, and A.~{Ribeiro}, ``Optimal graph-filter design
  and applications to distributed linear network operators,'' \emph{IEEE Trans.
  Signal Process.}, vol.~65, no.~15, pp. 4117--4131, Aug. 2017.

\bibitem{Gama:SPM20}
F.~{Gama}, E.~{Isufi}, G.~{Leus}, and A.~{Ribeiro}, ``Graphs, convolutions, and
  neural networks: {From} graph filters to graph neural networks,'' \emph{IEEE
  Signal Process. Mag.}, vol.~37, no.~6, pp. 128--138, Nov. 2020.

\bibitem{Isufi:TPAMI}
E.~{Isufi}, F.~{Gama}, and A.~{Ribeiro}, ``Edge{N}ets: {Edge} varying graph
  neural networks,'' \emph{IEEE Trans. Pattern Anal. Mach. Intell.}, 2022, to
  be published.

\end{thebibliography}

\end{document}